\documentclass[a4paper,11pt]{article}
\usepackage{caption}
\usepackage{jcappub} 
\usepackage{siunitx}
\usepackage{subcaption}

\title {Pulsational mode stability in complex EiBI-gravitating polarized astroclouds with \boldmath $(r,q)$-distributed electrons}

\author{Dipankar Ray\href{https://orcid.org/0009-0002-2308-7850}{\includegraphics[scale=0.1]{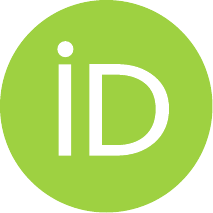}\hspace{1mm}}}\author[1]{and Pralay Kumar Karmakar{\href{https://orcid.org/0000-0002-3078-9247}{\includegraphics[scale=0.1]{orcid.pdf}\hspace{1mm}} \note{Corresponding author.}}}
\affiliation{Department of Physics, Tezpur University, \\ Napaam, Tezpur, 784028, Assam, India}

\emailAdd{dray056@tezu.ernet.in}
\emailAdd{pkk@tezu.ernet.in}

\abstract{The pulsational mode of gravitational collapse (PMGC) originating from the combined gravito-electrostatic interaction in complex dust molecular clouds (DMCs) is a canonical mechanism leading to the onset of astronomical structure formation dynamics. A generalized semi-analytic model is formulated to explore the effects of the Eddington-inspired Born-Infeld (EiBI) gravity, non-thermal $(r,q)$-distributed electrons, and dust-polarization force on the PMGC stability concurrently. The thermal ions are treated thermo-statistically with the Maxwellian distribution law and the non-thermal electrons with the  $(r,q)$-distribution law. The constitutive partially ionized dust grains are modeled in the fluid fabric. A spherical normal mode analysis yields a generalized linear PMGC dispersion relation. Its oscillatory and propagation characteristics are investigated in a judicious numerical platform. It is found that an increase in the polarization force and positive EiBI parameter significantly enhances the instability, causing the DMC collapse and vice versa. The electron non-thermality spectral parameters play as vital stabilizing factors, and so on. Its reliability and applicability are finally outlined in light of astronomical predictions previously reported in the literature.}
\begin{document}
\maketitle
\flushbottom
\section{Introduction}\label{sec:introduction}
Interstellar molecular clouds, vast swirling heterogeneous dust and gas mixtures, serve as the birthplaces for diverse bounded astrophysical structure evolution, such as stars, planets, and so forth \cite{Pandey_2002,Jeans:1902fpv,2009itss.book.....P,choudhuri2010astrophysics}. The fundamental physics of bounded structure formation has remained as a long-standing obscure research area yet to be well illuminated. In 1902, Sir James Hopwood Jeans, a renowned British astronomer and theoretical physicist, has introduced the Jeans (gravitational) instability criterion to elucidate the gravitational collapse dynamics in nebular gaseous media. According to his gravitational collapse theory \cite{Jeans:1902fpv}, the stellar structure formation process gets naturally initiated with the onset of gravitational instability. This instability gets triggered against the usual hydrostatic equilibrium configuration conjugationally established by the inward gravitational pressure force and the outward thermal pressure force. When the mass or length of a molecular cloud exceeds the critical Jeans mass-length scale limits, the gravitational pressure force dominates the thermal pressure force. It results in the collapse of giant molecular clouds fragmented into local cloudlets, consequently condensed in the form of stellarsimals, planetesimals, protostructures, and so forth \cite{chen1984introduction,choudhuri1998physics,eggleton2011chandrasekhar,hoyle1953fragmentation}.
\par
In self-gravitating complex (dusty) plasmas composed of electrons, ions, neutral, and charged (positive or negative) dust, the presence of charged dust grains generates a repulsive electrostatic force that opposes the inward gravitational force. Additionally, because of the greater mass of the dust particles against the lighter plasma compositions (electrons and ions), the self-gravitational effects of the dust species become significant in extensive astrophysical systems, like dust molecular clouds (DMCs). The DMCs are the dense sites of the interstellar medium (ISM), consisting of gas and micron to sub-micron-sized dust (silicate, graphite, magnetite, etc.) \cite{Pandey_2002,avinash2006dynamics}. The dust grains present in the partially ionized DMCs modify the traditional Jeans instability. Thus, in order to adapt the Jeans instability for structure formation in complex self-gravitating DMCs, a Jeans-like hybrid mode known as the pulsational mode of gravitational collapse (PMGC) has been introduced in 1999 by Dwivedi et al. \cite{dwivedi1999pulsational}. This hybrid instability emerges in a partially ionized DMC due to the interaction between gravitational and electrostatic forces, termed as gravito-electrostatic coupling \cite{Pandey_2002,dwivedi1999pulsational}.
\par 
Since the initial work on the PMGC instability, several researchers have shown a keen interest in this mode of gravitational collapse. Consequently, various scenarios in both linear and non-linear regimes have been explored. Extensive analyses of the PMGC dynamics in the presence of dust charge fluctuations have been conducted, demonstrating that dust charge fluctuations play a crucial role in the PMGC dynamics excitable in the DMCs \cite{Pandey_2002,devi2023nonlinear,karmakar2017pulsational}. Subsequently, the area of PMGC dynamics has been extended to the non-linear regime, and several non-linear model formalisms have accordingly been developed for different scenarios \cite{BORAH2015416,devi2023nonlinear}. Apart from the above, the effects of various parameters, such as viscoelasticity \cite{dutta2017viscoelastic}, turbulence \cite{karmakar2017pulsational}, ion drag \cite{haloi2015nonlinear,yang2020effect}, and non-thermality \cite{bhakta2017pulsational,karmakar2017evolutionary,karmakar2024athermal} on the PMGC dynamics have been investigated. Recently, this field has further been expanded to modified gravity frameworks, such as
the EiBI gravity \cite{das2024dynamics}, zeroth-order gradient effects \cite{borah2016atypical}, and so forth.
\par
It is seen in the literature that the impact of polarization force on the PMGC dynamics, which is a critical aspect of inhomogeneous plasmas, has yet to be extensively addressed. In non-uniform plasmas with uneven distributions of electrons and ions, characterized by the absence of dust charge fluctuations, the deformation of the plasma Debye sheaths gives rise to a new force known as the polarization force ($\Vec{F}_p$). It is in addition to the conventional Coulomb electrostatic force ($\Vec{F}_e=q_d \Vec{E}$; where, $q_d$ and $\Vec{E}$ stand for the dust charge and electric field, respectively) \cite{hamaguchi1994polarization}. This sheath-polarization force acting on the charged dust particles vanishes entirely in uniform plasma environments. In complex plasmas comprising of negatively charged dust grains, the Debye sheath is formed by ions around the dust grains. The distortion of such plasma sheaths results in the development of polarization force \cite{el2020effects}.
\par
The expression for the polarization force is given as $\Vec{F}_p= -q_d^2\Vec{\nabla}\lambda_D /(2\lambda_D^2)$; where, $\lambda_D = \lambda_{Di}/\sqrt{1+ (\lambda_{Di}/\lambda_{De})^2}$ denotes the effective Debye length \cite{dolai2020effects,khrapak2009influence}. Here, $\lambda_{De(i)} = \sqrt{T_{e(i)} / 4\pi e^2 n_{e(i)}}$ represents the electron (ion) Debye length, with $T_{e(i)}$ being the electron (ion) temperature, $e$ is the elementary charge (electronic), and $n_{e(i)}$ corresponds to the number density of electrons (ions). By utilizing the approximation $T_i n_e \ll T_e n_i$ \cite{ashrafi2014polarization}, which is suitable for negatively charged dust, and $T_i 
\Vec{\nabla} n_i=-n_i e \Vec{\nabla} \phi$ \cite{khrapak2009influence}, a simplified form for the polarization force can be derived as $\Vec{F}_p = - |q_d| R (n_i/n_{i0})^{1/2}\Vec{\nabla} \phi$ \cite{ashrafi2014polarization} (refer to appendix \ref{sec:appendix} for details). Here, $R = |q_d| e/(4T_i\lambda_{Di0})$ serves as the polarization interaction parameter that determines the influence of the polarization force, with $n_{i0}$ and $\phi$ representing the equilibrium ion number density and electrostatic potential, respectively. It is evident from here that a higher dust charge (or lower ion temperature) results in a stronger polarization force and vice versa.
\par
Numerous studies in the literature have relied on the Maxwellian velocity distribution of the lighter constitutive species to elucidate plasma dynamics \cite{dwivedi1999pulsational,karmakar2011nonlinear,yang2020effect,das2024dynamics}. However, advancements in observational capabilities have unveiled that the particle distribution of space plasmas significantly deviates from the Maxwellian velocity distribution. In 1983, flat-top characteristics in the electron velocity distributions at the boundary of the Earth's bow shock have been reported \cite{feldman1983electron}. Multiple spacecraft missions in the downstream region of the bow shock have subsequently observed this behaviour of low-energy electrons. The ARTEMIS spacecraft has provided further evidence for this phenomenon in the geomagnetic tail region \cite{zhao2017electron}. Additionally, similar flat-top distribution features near the magnetic reconnection regions have been reported by analyzing data collected from the four Cluster satellites \cite{asano2008electron}. Observational studies of such space plasmas have established the widespread presence of particle populations characterized by high or super-thermal energy tails that do not adhere to the standard Maxwellian velocity distribution. These non-Maxwellian (non-equilibrium) distributions exhibit an approximate power-law dependence in velocity space, implying a substantial fraction of particles with energies significantly greater than the average thermal energy value. Examples in this context are the interstellar medium \cite{livadiotis2011first}, solar wind \cite{gosling1981interplanetary}, planetary magnetospheres \cite{krimigis1983general,vasyliunas1968survey}, planetary nebulae \cite{zhang2004electron,Nicholls_2012}, H \textsc{ii} regions \cite{Nicholls_2012}, and so forth. 
\par
To model the non-Maxwellian behaviours, various non-thermal velocity distribution laws have been proposed, such as the kappa distribution \cite{vasyliunas1968survey}, $q$-nonextensive distribution \cite{tsallis1988possible}, Cairns distribution \cite{cairns1995electrostatic}, and so on. In this context, the kappa distribution law has been demonstrated to be effective in characterizing the high-energy particle tails. Likewise, the kappa distribution is also characterized by a single non-thermal spectral index $(\kappa)$, which typically controls the high-energy tails. To explain both the flat-top and high-energy tail features simultaneously, Qureshi et al. \cite{qureshi2004parallel} have introduced a new bi-spectral non-Maxwellian electron velocity distribution function called the generalized $(r,q)$-distribution function in 2004. The $(r,q)$-distribution law specifies two non-thermal spectral indices ($r$ and $q$) that regulate the electron population in different parts of the energy spectrum. Specifically, $r$ controls the electron population at the flat-tops, while $q$ governs the high-energy tails \cite{zaidi2024effect,qureshi2004parallel}. The $(r,q)$-distribution is a generalized form of the kappa distribution, which can be reduced to the kappa distribution (for $r=0$ and $q\to \kappa +1$) as well as the Maxwellian distribution (for $r=0$ and $q\to \infty$). Therefore, the $(r,q)$-distribution thermo-statistically gives more flexibility in stability analyses as compared with other non-thermal velocity distribution laws.
\par
Einstein's general theory of relativity (GTR) has been proven to be one of the most successful theories explaining gravity. After the introduction of GTR in 1915, Eddington was the very first person to experimentally verify GTR in 1919 by measuring the bending of light during a solar eclipse \cite{dyson1923determination}. Since then, GTR has passed numerous tests (in weak-field regime) in diverse astrophysical environments, such as the solar system \cite{shapiro1990solar,will2014confrontation}, with binary pulsars \cite{stairs2003testing}, black hole binaries \cite{abbott2016binary,abbott2016observation}, and in proximity to the Galactic Centre \cite{hees2017testing}. Despite the remarkable accomplishments of GTR, it falls short in tackling certain inherent challenges. For instance, it fails to explain the rapid cosmic expansion of the universe, the formation of singularities in the space-time fabric during the gravitational collapse of stars and in the early universe \cite{banerjee2017constraints, banerjee2022stellar}, as well as the enigma surrounding dark energy and dark matter \cite{jana2018constraints}. To overcome these critical issues, many extensions of GTR have been proposed, such as Scalar-Tensor-Vector gravity (STVG) \cite{moffat2006scalar}, 4D Einstein-Gauss-bonnet gravity (4DEGB) \cite{glavan2020einstein,zanoletti2024cosmological}, $f(R)$ gravity \cite{starobinsky1980new}, Eddington inspired Born-Infeld (EiBI) gravity \cite{banados2010eddington}, etc. After the seminal work on the Jeans instability of self-gravitating objects in non-relativistic Newtonian gravity \cite{Jeans:1902fpv}, it has been further extended to GTR \cite{lifshitz1946gravitational}. Recently, the impact of different alternative gravity formalisms on the Jeans collapse dynamics has been thoroughly investigated \cite{yang2023jeans,yang2020jeans,vainio2016jeans,he2022jeans,hazarika2024polytropic}.
\par 
Among the various extensions, the EiBI gravity has been successful in demonstrating the elimination of the singularities in both the primordial universe and subsequent gravitational collapse scenarios \cite{banados2010eddington,pani2012eddington}. The development of EiBI gravity has been motivated by the Born–Infeld action for nonlinear electrodynamics and it has been formulated using Eddington's theory of gravity \cite{yang2023jeans}. Introduced by Ba\~{n}ados and
Ferreira in 2010, the EiBI gravity theory has been scrutinized in several astrophysical scenarios, such as the solar system \cite{casanellas2011testing}, compact objects \cite{banerjee2017constraints,pani2011compact}, and even in cosmological scales \cite{harko2014bianchi,avelino2012eddington}; hence, it can be treated as a reliable extension of GTR. The extensions of GTR are generally obtained through modifications to the Einstein-Hilbert action expressed as (in the unit of $c$) $S_\text{EH}=1/(16\pi G)\int d^4x \sqrt{-g}(R-2\Lambda)$ \cite{banerjee2022stellar,banerjee2017constraints,clifton2012modified}. Here, $g$ represents the determinant of the metric $g_{\mu\nu}$, $R$ is the Ricci scalar defined as $R=g^{\mu\nu}R_{\mu\nu}$, $G$ is the universal gravitational constant, and $\Lambda$ denotes the cosmological constant. The famous Einstein's field equations can be obtained by varying the Einstein-Hilbert action with respect to the metric $g_{\mu\nu}$ as $R_{\mu\nu}-(1/2) g_{\mu\nu}R + \Lambda g_{\mu\nu} = 8\pi G T_{\mu\nu}$ \cite{banados2010eddington,clifton2012modified,schutz2022first}; where, $R_{\mu\nu}$ is the Ricci Tensor and $T^{\mu\nu}=(g)^{-1/2}\partial S_M/\partial g_{\mu\nu}$ denotes the stress-energy tensor with $S_M$ being the matter action.
\par 
It is noteworthy that following the remarkable works by Deser et al. \cite{deser1998born} and Vollick \cite{vollick2004palatini,vollick2005born}, Ba\~{n}ados and Ferreira have considered a Born-Infled type action of the form $S_\text{EiBI}=1/(8\pi \chi G)\int d^4x \left(\sqrt{|g_{\mu\nu}+\chi R_{\mu\nu}}|-\lambda\sqrt{-g} \right)+S_M (g_{\mu\nu},\psi_M)$ \cite{banados2010eddington,banerjee2022stellar,banerjee2017constraints}. Here, $\chi$ serves as the independent parameter for EiBI gravity, $\lambda (\ne 0)$ is a dimensionless constant, and $\psi_M$ represents the matter field. It is important to note here that in the limit where $\chi R \ll 1$ and $\lambda=(\Lambda\chi+1)$, the EiBI action reduces to the conventional Einstein-Hilbert action \cite{banerjee2017constraints}. The field equations of EiBI gravity can be obtained from the EiBI action $(S_\text{EiBI})$ using the Palatini formalism by introducing an auxiliary metric $q_{\mu\nu}=g_{\mu\nu}+\chi R$, leading to $R_{\mu\nu} \approx \Lambda g_{\mu\nu}+8\pi G \left(T_{\mu\nu}-(1/2)g_{\mu\nu}T\right)+8\pi G \chi \left(S_{\mu\nu}-(1/4)Sg_{\mu\nu}\right)+\mathcal{O}(\chi^2)$ \cite{pani2012surface,banerjee2017constraints}. Here, $S_{\mu\nu}=T_{\mu\nu}-(1/2)TT_{\mu\nu}$ is the source tensor, while $S$ and $T$ represent the traces of $S_{\mu\nu}$ and $T_{\mu\nu}$, respectively. The third term on the right side of the EiBI-modified field equations introduces corrections to the standard Einstein's field equations. As $\chi$ approaches zero, the familiar Einstein's field equations are recovered.
\par
It is clear from the above scenarios that the EiBI gravity, non-thermal $(r,q)$-distributed electrons, and dust-polarization force play significant roles in the collapse of a molecular cloud. Motivated by this, a generalized semi-analytic model is proposed in this paper incorporating the aforementioned factors to study the PMGC stability dynamics of a complex DMC.
\par
After the introduction given in section \ref{sec:introduction}, the rest of the paper is organized as follows. Section \ref{sec:Physical model and mathematical formalism}, outlines the physical model formalism and the basic governing equations describing the system. Section \ref{sec:Linearization and Fourier analysis}, explains the mathematical method for linearization of the governing equations followed by the spherical Fourier analysis. A generalized linear dispersion relation (quartic in degree) is obtained using the method of decoupling (elimination). In section \ref{sec:Results and Discussions}, the numerical outcomes, along with figures, are discussed. Finally, the results are concluded with
a brief summary and future scope of this investigation in section \ref{sec:Conclusions}.
\section{Physical model and mathematical formalism}\label{sec:Physical model and mathematical formalism}
To study the linear PMGC dynamics, a weakly coupled $(\Gamma_\text{Cou}\ll 1)$, inhomogeneous, and globally quasi-neutral self-gravitating DMC is considered, ignoring the effects of rotation, magnetic field, and tidal action resulting from the gravitational effects of distant astrophysical objects. It consists of four species, namely, $(r,q)$-distributed (non-thermal) electrons, Maxwellian (thermal) ions, neutral and negatively charged dust particles of identical size $(r_d)$ and mass $(m_d)$. The Coulomb coupling parameter $(\Gamma_\text{Cou})$ associated with the system refers to the ratio of dust potential energy to dust thermal energy and is mathematically expressed as $\Gamma_\text{Cou}=(1/4\pi \epsilon_0)\left(q_d^2/(a_d T_d)\right)\left(\text{in cgs, }\Gamma_\text{Cou}= q_d^2/(a_d T_d)\right)$; where, $a_d = \left(3/(4\pi n_{d0})\right)^{1/3}$ denotes the Wigner-Seitz radius of the dust grains and $T_d$ represents the dust temperature (in eV) \cite{kalita2021jeans}. A system is said to be weakly coupled for $\Gamma_\text{Cou}\ll1$ and strongly coupled for $\Gamma_\text{Cou}\gg1$ \cite{Shukla_2002}. The considered weakly coupled DMCs are indeed partially ionized in nature with an estimated degree of ionization $\sim 10^{-7}$ \cite{shu1987star}. Owing to their low ionization level, both neutral and charged dust particles coexist with electrons and ions in such DMCs. A further simplification is considered by ignoring the dust charge fluctuations which is justifiable when the hydrodynamic time scale $(\tau_{d})$ significantly exceeds the grain charging time scale $(\tau_{c})$, i.e., $\tau_d \gg \tau_c$ \cite{gupta2001effect}. Here, $\tau_{d} = \omega_{pd}^{-1}$, $\omega_{pd}$ being the dust plasma frequency and $\tau_{c} = \nu_{d}^{-1}$, $\nu_{d}$ denotes the grain charging frequency. In a typical DMC, the hydrodynamic time scale is approximately $\tau_d \sim 10^{-2}$ seconds, while the dust charging time scale is $\tau_c \sim 10^{-8}$ seconds \cite{jian2009effects}. Therefore, dust charge fluctuations can be safely disregarded for this particular DMC.
\begin{figure}[htbp]
    \centering
    \includegraphics[width=1\textwidth] {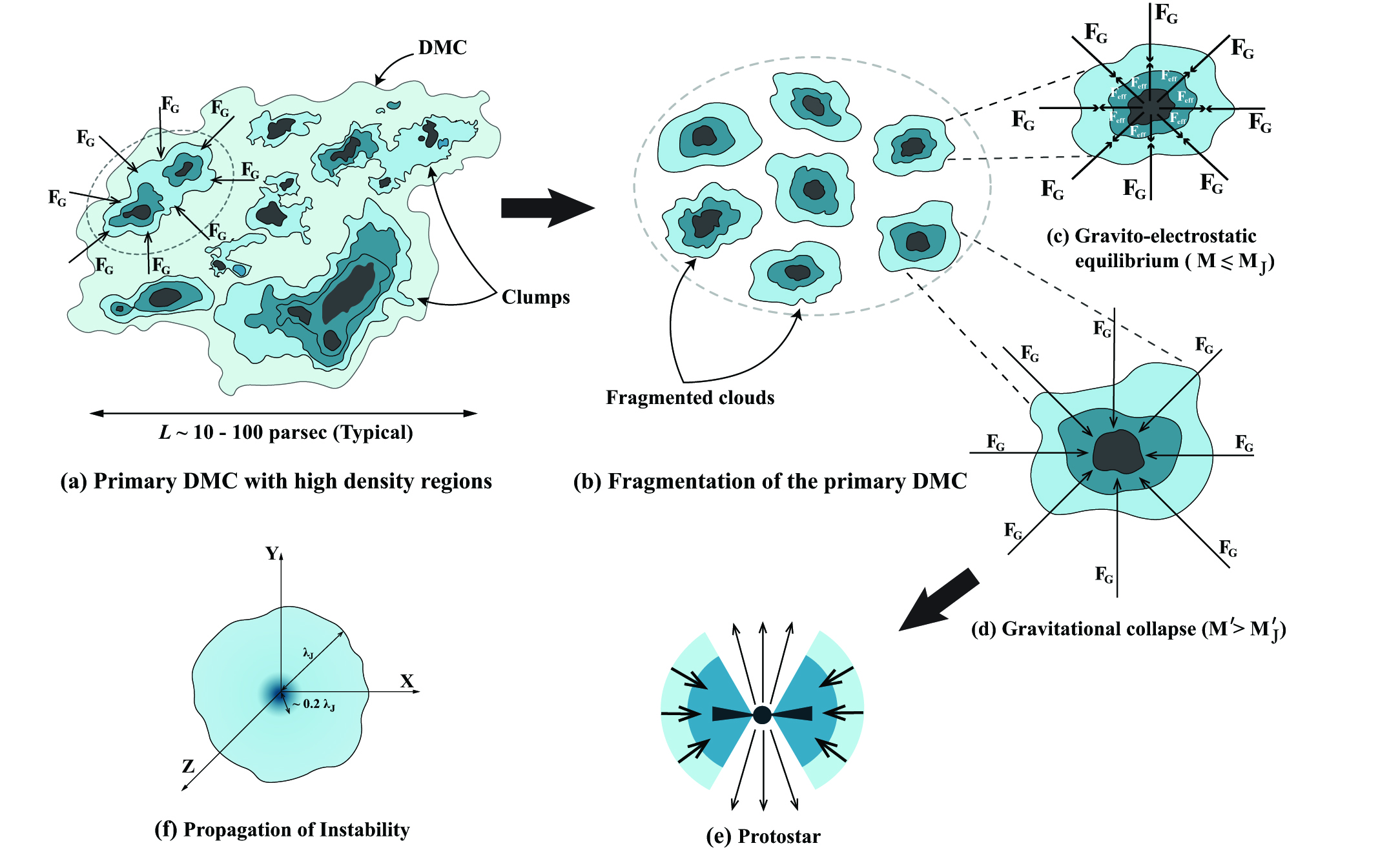}
    \caption{A schematic showing the fragmentation processes of DMCs leading to stellar structure formation triggered by the PMGC instability.}
    \label{fig:figure1}
\end{figure}
\par
The DMC fragmentation process due to PMGC instability can be illustrated with a schematic diagram as in figure \ref{fig:figure1}. Here, the complex DMC (figure \ref{fig:figure1}a) is subjected to various external perturbations, such as explosive shock waves propagating from nearby supernova explosions, cloud-cloud collisional interactions, etc. These perturbations may produce some highly dense regions compared to the overall density of the DMC in the form of diverse parametric inhomogeneities (gradient effects). Consequently, these high-density regions start to contract due to the inward self-gravitational pull among the constitutive heavier dust particles, leading to a rise in outward thermal pressure. However, due to the very low initial density of the cloud, the thermal pressure can escape from the parent cloud in the form of various deceptive processes, such as thermal conduction, radiation, etc. \cite{ward2011introduction,2009itss.book.....P}. Hence, the outward thermal pressure does not balance the inward gravitational pressure at the initial stage of collapse. In contrast, the effective electrostatic repulsion (electrostatic force between the identical charged particles and polarization force) tries to counter the gravitational collapse of the cloud. Depending upon the critical Jeans mass-length limit, a gravito-electrostatic equilibrium is established, forming a stable cloud region (figure \ref{fig:figure1}c). However, if the mass of the cloud exceeds the Jeans critical mass limit, the self-gravitational force prevails over the electrostatic repulsion force. It results in a growing instability in the cloud (figure \ref{fig:figure1}f), leading to the fragmentation of the original cloud into (smaller) cloudlets (figure \ref{fig:figure1}b). These equilibrium and collapse processes continue till the cloud is fragmented to a smaller stellar mass scale, resulting in a protostar formation (figure \ref{fig:figure1}e). As the cloud density and temperature increase with each condensation stage, they eventually become opaque to the thermal radiation, and the radiation can no longer escape from the stellar body, eventually establishing a hydrostatic equilibrium \cite{ward2011introduction}.
\par
The electrons are assumed to follow the generalized $(r,q)$ thermo-statistical distribution law given in generic notations \cite{qureshi2004parallel,ali2021contribution} as 
\begin{equation}
    \label{eq:2.1}
    F_{r,q}(v)= \frac{\alpha}{\pi v_\text{th}^{3}}\left(1+\frac{1}{q-1}\left[\frac{v^2-2e\phi/m_e}{\beta\left(2T_e/m_e\right)}\right]^{(r+1)}\right)^{-q};
\end{equation} 
\begin{equation}
    \label{eq:2.2}
    \alpha=\frac{3\Gamma[q](q-1)^{-\frac{3}{(2+2r)}}}{4\beta^{\frac{3}{2}}\Gamma\left[q-\frac{3}{2+2r}\right]\Gamma\left[1+\frac{3}{2+2r}\right]},
\end{equation}
\begin{equation}
    \label{eq:2.3}
    \beta = \frac{3(q-1)^{-\frac{1}{(1+r)}}\Gamma\left[q-\frac{3}{2+2r}\right]\Gamma\left[\frac{3}{2+2r}\right]}{2\Gamma\left[q-\frac{5}{2+2r}\right]\Gamma\left[\frac{5}{2+2r}\right]}.
\end{equation}
The number density of non-thermal $(r,q)$-distributed electrons $(n_e)$ can be obtained by integrating  \eqref{eq:2.1} over a velocity space as  \cite{ali2021contribution}
\begin{equation}
    \label{eq:2.4}
    n_e=n_{e0}\left[1+A\frac{e\phi}{T_e}+B\left(\frac{e\phi}{T_e}\right)^2\right];
\end{equation}

\begin{equation}
    \label{eq:2.5}
     A=\frac{(q-1)^{-\frac{1}{(1+r)}}\Gamma\left[q-\frac{1}{2+2r}\right]\Gamma\left[\frac{1}{2+2r}\right]}{2\beta\Gamma\left[q-\frac{3}{2+2r}\right]\Gamma\left[\frac{3}{2+2r}\right]},
\end{equation}

\begin{equation}
    \label{eq:2.6}
    B=-\frac{(1+4r)(q-1)^{\frac{-2}{(1+r)}}\Gamma \left[q+\frac{1}{(2+2r)}\right]\Gamma \left[-\frac{1}{2+2r} \right]}{8\beta^2\Gamma \left[q-\frac{3}{2+2r} \right] \Gamma \left[\frac{3}{2+2r} \right]}.
\end{equation}
Due to the heavier mass of the ions than that of the electrons, the inertia of the ions would be larger than that of the electrons. Thus, the ions are considered to follow the Maxwellian velocity distribution on a slow dust inertial time scale, with its number density given as \cite{karmakar2011nonlinear}
\begin{equation}
    \label{eq:2.7}
    n_i=n_{i0}\exp\left(-\frac{e\phi}{T_i}
    \right).
\end{equation}
In equations \eqref{eq:2.1}-\eqref{eq:2.7}, $\Gamma [z] \left(\text{defined simply as } \Gamma [z] =\int_{0}^{\infty} e^{-x} x^{z-1} dx\right)$ denotes the so-called characteristic gamma function \cite{qureshi2004parallel}. The symbols $r$ and $q$ represent the non-thermal spectral indices specifying the electronic thermo-statistics. The variable $\phi$ represents the electrostatic potential, $v_\text{th}=(2T_e/m_e)^{1/2}$ denotes the thermal velocity of the electrons, $T_{e(i)}$ corresponds to the electron (ion) temperature (in eV), $m_e$ is the mass of an electron, and $n_{e(i)}$ is the number density of electrons (ions), with $n_{e0(i0)}$ being their corresponding equilibrium number densities. It is crucial to emphasize that the spectral indices $r$ and $q$ in the $(r,q)$-distribution law must follow the conditions $q>1$ and $q(r+1)>5/2$ \cite{qureshi2004parallel,ali2021contribution}. In the special case when $r=0$ and $q \to \infty$, the $(r,q)$-distribution transitions to the Maxwellian distribution; where, the multi-order non-thermality coefficients A and B become $A=1$ and $B=1/2$. 
\par
The dust particles, when emerged in a plasma environment, become electrically charged (through contact electrification process) due to the random collisions of lighter constituent species with the dust grains. As the electrons possess a greater thermal velocity compared to the ions, they reach the dust grain surface significantly faster than the ions. This rapid accumulation of the electrons on the grain surfaces makes them negatively charged \cite{Shukla_2002,dutta2016stability}. However, due to the partially ionized state of the cloud, not all dust grains become electrically charged. This leads to the presence of both neutral and negative dust grains in the system. Thus, the neutral and negatively charged grains are treated as two distinct fluids. Furthermore, in the context of studying the collapse of a dense DMC, it is relevant to consider the momentum exchange between charged and neutral dust grains. The rates of binary collisions transferring momentum from charged to neutral dust particles and vice versa are respectively given as $\nu_{cn}\approx \pi r_d^2 n_{dn0} v_\text{td}$ and $\nu_{nc}\approx \pi r_d^2 n_{dc0} v_\text{td}$. Here, $r_d$ is the dust radius, $n_{dco} (n_{dn0})$ is the equilibrium number density of charged (neutral) dust, and $v_\text{td}=(2T_d/m_d)^{1/2}$ is their common thermal velocity. The frictional force term can be neglected for neutral dust under the condition that the Jeans mode frequency $\omega\gg\nu_{cn}$ with $\nu_{cn}/\nu_{nc}=n_{dn0}/n_{dc0}\gg1$ \cite{dwivedi1999pulsational}.
\par
The neutral dust dynamics can be described using the continuity and momentum equations with the negligible frictional effect respectively as
\begin{equation}
    \label{eq:2.8}
    \partial_t n_{dn}+ \Vec{\nabla}.(n_{dn}\Vec{v}_{dn})=0,
\end{equation}
\begin{equation}
   \label{eq:2.9} \partial_t \Vec{v}_{dn}+(\Vec{v}_{dn}.\Vec{\nabla}) \Vec{v}_{dn}=-\Vec{\nabla}\psi.
\end{equation}
Again, the charged dust dynamics can be described using continuity and momentum equations with significant frictional effect respectively as 
\begin{equation}
    \label{eq:2.10}
    \partial_t n_{dc}+\Vec{\nabla}.(n_{dc}\Vec{v}_{dc})=0,
\end{equation}
\begin{equation}
    \label{eq:2.11}
    \partial_t \Vec{v}_{dc}+(\Vec{v}_{dc}.\Vec{\nabla})\Vec{v}_{dc}=-\frac{q_d}{m_d}\Vec{\nabla}\phi+\frac{q_d}{m_d}R\left(\frac{n_i}{n_{i0}}\right)^{1/2}\Vec{\nabla}\phi-\Vec{\nabla}\psi-\nu_{cn}(\Vec{v}_{dc}-\Vec{v}_{dn}).
\end{equation}
In the above equations \eqref{eq:2.8}-\eqref{eq:2.11}, $n_{dn(dc)}$ and $\Vec{v}_{dn(dc)}$ stand for the neutral (charged) dust number density and neutral (charged) dust fluid velocity, respectively. Here, $q_d = -Z_de$ denotes the dust charge with $Z_d=|q_d/e|$ being the dust charge number and $e$ is the elementary charge (electronic). Moreover, $R$ represents the polarization interaction parameter, and $\nu_{cn}$ is the binary collisional rate of momentum transfer from the charged to the neutral dust grains. Furthermore, $\phi$ and $\psi$ respectively denote the electrostatic and gravitational potential.
\par
The modified Poisson equation serves as the starting point for phenomenological studies using EiBI gravity. The required Poisson equation can be obtained from the modified Einstein's field equation (mentioned in section \ref{sec:introduction}) by taking the weak field limit as \cite{banerjee2017constraints,avelino2012eddington}
\begin{equation}
    \label{eq:2.12}
     \nabla^2\psi = 4\pi G \rho_{m_j} + \left(\frac{\chi}{4}\right)\nabla^2 \rho_{m_j},
\end{equation}
 where, $\psi$ represents the gravitational potential, $\rho_{m_j}=n_j m_j \ (j=e,i,dc,dn)$ denotes the matter density, and $\chi$ serves as the EiBI parameter. Numerous constraints have been imposed on the EiBI parameter for different astrophysical objects, such as white dwarfs $(\chi \le 4.86 \times 10^{3}$ \unit{kg^{-1}.m^5.s^{-2})} \cite{banerjee2017constraints}, neutron stars $(\chi<10^{-2}$ \unit{kg^{-1}.m^5.s^{-2})} \cite{pani2012eddington,pani2011compact}, Sun $(\chi \le 3 \times 10^{5}$ \unit{kg^{-1}.m^5.s^{-2})}\cite{casanellas2011testing}, and so forth. The tightest nuclear constraint on the EiBI parameter $(\chi)$ has been imposed by Avelino in 2012 \cite{avelino2012eddington}, given as $\chi \le 10^{-3}$ \unit{kg^{-1}.m^5.s^{-2})}. It is important to note that the usual gravitational Poisson equation can be recovered from \eqref{eq:2.12} for $\chi =0$.
\par
Finally, the DMC model is closed by utilizing the electro-gravitational Poisson equations, which provide the potential distributions generated by the respective density fields, cast as
\begin{equation}
    \label{eq:2.13}
    \nabla^2\phi=4\pi e(n_e-n_i-\frac{q_dn_{dc}}{e}),
\end{equation}
\begin{equation}
   \label{eq:2.14}
   \nabla^2\psi=4\pi Gm_d(n_{dc}+n_{dn}-n_{d0})+\left(\frac{\chi}{4}\right)\nabla^2[m_d(n_{dc}+n_{dn}-n_{d0})].
\end{equation}
Here, the first term on the right side of \eqref{eq:2.14} signifies the effect of Newtonian gravity, whereas the second term accounts for the modification made by the EiBI gravity, and $n_{d0}(=n_{dc0}+n_{dn0})$ is the net equilibrium dust number density, serving as the Jeans swindle.
 \par 
As the astrophysical self-gravitating plasmas are usually confined in spherical geometric volumes, a spherically symmetric approximation (radial 1-D case) is employed rather than a full spherical geometry (spherical 3-D case). This simplification is justifiable under the assumption of radial symmetry, which holds when the fluctuation wavelength significantly exceeds the grain-to-grain distance \cite{karmakar2011nonlinear}. The fundamental governing equations \eqref{eq:2.8}-\eqref{eq:2.14} defining the DMC dynamics can be written in the spherically symmetric coordinate $(\rho,t)$ with all usual notations as
\begin{equation}
   \label{eq:2.15} \partial_tn_{dn}+\rho^{-2}\partial_\rho\left(\rho^2n_{dn}v_{dn}\right)=0,
\end{equation}
\begin{equation}
   \label{eq:2.16} \partial_tv_{dn}+(v_{dn}\partial_\rho)v_{dn}=-\partial_\rho\psi,
\end{equation}
\begin{equation}
   \label{eq:2.17} \partial_tn_{dc}+\rho^{-2}\partial_\rho\left(\rho^2n_{dc}v_{dc}\right)=0,
\end{equation}
\begin{equation}
   \label{eq:2.18} \partial_tv_{dc}+\left(v_{dc}\partial_\rho\right)v_{dc}=-\frac{q_d}{m_d}\partial_\rho \phi+\frac{q_d}{m_d}R\left(\frac{n_i}{n_{i0}}\right)^{\frac{1}{2}}\partial_\rho \phi-\partial_\rho \psi-\nu_{cn}\left(v_{dc}-v_{dn}\right),
\end{equation}
\begin{equation}
   \label{eq:2.19} \rho^{-2}\partial_\rho \left(\rho^2\partial_\rho\phi\right) = 4 \pi e \left(n_e-n_i-\frac{q_dn_{dc}}{e}\right),
\end{equation}
\begin{equation}
    \label{eq:2.20}
    \begin{aligned}
        \rho^{-2}\partial_\rho\left(\rho^2\partial_\rho\psi\right) &= 4\pi Gm_d\left(n_{dc}+n_{dn}-n_{d0}\right) \\
&\quad + \left(\frac{\chi}{4}\right)\rho^{-2}\partial_\rho\left(\rho^2\partial_\rho\left[m_d(n_{dc}+n_{dn}-n_{d0})\right]\right).
    \end{aligned}
\end{equation}
Equations \eqref{eq:2.15}-\eqref{eq:2.20} are the governing equations required to describe the dynamics of the spherical DMC model. Here, $\partial_\rho \equiv \partial/\partial \rho $ and $\partial_t \equiv \partial/\partial t $ signify the space and time gradient operators, respectively. It is noteworthy here that the geometrical curvature effects are responsible for the formation of $1/\rho$ terms, and the planar equations can be retraced under the geometric approximation $\rho\to\infty$, $\rho$ being the radial distance.

\section{Linearization and Fourier analysis}\label{sec:Linearization and Fourier analysis}
Employing a standard approach of spherical wave analysis \cite{dasgupta2019jeans,kalita2021jeans,hazarika2024polytropic}, a linear perturbation $(f_1)$ is introduced to the relevant physical fluid parameters around their respective homogeneous equilibrium values $(f_0)$. The multi-parametric perturbations with amplitude factor $f_{10}$ assume the mathematical shape of a symmetric spherical wave in a usual symbolism without any dimensional disparity in the absence of polar and azimuthal contributions as
\begin{equation}
    \label{eq:3.1}
    f(\rho,t) = (f_0 + f_1) = f_0 + \rho^{-1}f_{10}\exp \left[-i(\omega t - k\rho)\right];
\end{equation}
\begin{equation}
    \label{eq:3.2}
    f=[n_{dn}\;n_{dc}\;n_e\;n_i\;v_{dn}\;v_{dc}\;\phi\;\psi]^T,
\end{equation}
\begin{equation}
    \label{eq:3.3}
    f_0=[n_{dn0}\;n_{dc0}\;n_{e0}\;n_{i0}\;0\;0\;0\;0]^T,
\end{equation}
\begin{equation}
    \label{eq:3.4}
     f_{1}=[n_{dn1}\;n_{dc1}\;n_{e1}\;n_{i1}\;v_{dn1}\;v_{dc1}\;\phi_1\;\psi_1]^T.
\end{equation}
As already mentioned above, it is repeated broadly that the symbol, $f_{1}$, represents a small (linear) perturbation of $f$ about $f_0$ such that $f_1 \approx \rho^{-1}f_{10}\exp\left(-i(\omega t-k\rho)\right)$. This sinusoidal spherical wave has angular frequency $\omega$ and angular wavenumber $k$.
It is noteworthy that $f_0$ and $\rho^{-1}f_{10}$ have the same dimensions. $T$ represents the transpose operation thereof. The exponential term embodies the inherent wave-like nature of the perturbations. The inherent nature of the adopted linearization approach dictates the exclusion of all higher-order terms, as their contributions become insignificant in the current model analysis.
The linear differential operators accordingly transform as $\partial_\rho \rightarrow (ik-\rho^{-1})$, $\partial_t \rightarrow (-i\omega)$, and $\partial_\rho^2 \rightarrow\left((-k^2+2\rho^{-2})-i(2k\rho^{-1})\right)$ \cite{kalita2021jeans}. The Fourier analysis is used to transform the fluctuation dynamics from the real coordination (direct) space $(\rho,t)$ to the wave (reciprocal) space $(k,\omega)$ in a physically judicious manner. Accordingly, the applied spherical Fourier analysis reduces the equations \eqref{eq:2.1},\eqref{eq:2.2}, and \eqref{eq:2.9}-\eqref{eq:2.14} in their linearized forms given respectively as 
\begin{equation}
    \label{eq:3.5}
    n_{e1}=n_{e0}\left(\frac{Ae\phi_1}{T_e}\right),
\end{equation}
\begin{equation}
    \label{eq:3.6}
    n_{i1}=-n_{i0}\left(\frac{e\phi_1}{T_i} \right),
\end{equation}
\begin{equation}
    \label{eq:3.7}
    n_{dn1}=\frac{n_{dn0}(k^2+\rho^{-2})}{\omega^2} \psi_1,
\end{equation}
\begin{equation}
    \label{eq:3.8}
    v_{dn1}=\left(\frac{ik-\rho^{-1}}{i\omega}\right) \psi_1, 
\end{equation}
\begin{equation}
    \label{eq:3.9}
    n_{dc1}=\frac{n_{dc0}(k^2+\rho^{-2})}{\omega^2}  \left[\psi_1+\left(\frac{q_d}{m_d} \right)(1-R) \left(1+\frac{i\nu_{cn}}{\omega}\right)^{-1} \phi_1 \right],
\end{equation}
\begin{equation}
    \label{eq:3.10}
    v_{dc1}=\frac{(q_d/m_d)(1-R)(ik-\rho^{-1})\phi_1+(ik-\rho^{-1})\psi_1-\nu_{cn}v_{dn1}}{i\omega -\nu_{cn}},
\end{equation}
\begin{equation}
    \label{eq:3.11}
    \phi_1=\frac{4\pi e}{k^2} \left[n_{i1}-n_{e1}+\left(\frac{q_d}{e}\right) n_{dc1} \right],
\end{equation}
\begin{equation}
    \label{eq:3.12}
    \psi_1=-\left(\frac{4\pi G}{k^2}-\frac{\chi}{4}\right)m_d(n_{dc1}+n_{dn1}).
\end{equation}
Substituting the value of $n_{dn1}$ from \eqref{eq:3.7} in \eqref{eq:3.12}, the perturbed gravitational potential $(\psi_1)$ can be obtained as follows
\begin{equation}
    \label{eq:3.13}
    \psi_1=-\left(\frac{4\pi G}{k^2}-\frac{\chi}{4}\right)m_d n_{dc1}\left[\frac{4k^2\omega^2}{4k^2 \omega^2+(k^2+\rho^{-2})\left(4\omega_J^2-4\omega_{Jc}^2 - \chi m_d n_{dn0}k^2\right)}\right].
\end{equation}
Here, $\omega_J = \sqrt{4 \pi G m_d n_{d0}}$ and $\omega_{Jc} = \sqrt{4 \pi G m_d n_{dc0}}$ are the effective critical Jeans frequency for the entire dust species and the critical Jeans frequency for the charged dust particles \cite{Jeans:1902fpv,karmakar2017evolutionary}.
Utilizing the value of perturbed gravitational potential $(\psi_1)$ from \eqref{eq:3.13} in \eqref{eq:3.9}, the perturbed charged dust number density $(n_{dc1}$) can be rewritten as 
\begin{equation}
    \label{eq:3.14}
    \begin{aligned}
        n_{dc1} &= \frac{n_{dc0}(k^2+\rho^{-2})}{\omega^2} \left[ \left(\frac{q_d}{m_d} \right)(1-R) \left(1+\frac{i\nu_{cn}}{\omega}\right)^{-1} \right] \\
        & \quad \times \left[ \frac{4 k^2 \omega^2 + (k^2+\rho^{-2}) \left(4 \omega_J^2-4\omega_{Jc}^2- \chi m_d n_{dn0}k^2 \right)}{4 k^2 \omega^2 + (k^2+\rho^{-2}) \left(4 \omega_J^2 - \chi m_d n_{d0}k^2 \right)} \right]\phi_1.
    \end{aligned}
\end{equation}
Finally, substituting the values of $n_{e1}$, $n_{i1}$, and $n_{dc1}$ from \eqref{eq:3.5},\eqref{eq:3.6}, and \eqref{eq:3.14} in \eqref{eq:3.11}, an unnormalized quartic dispersion relation is achieved as follows
\begin{equation}
    \label{eq:3.15}
    \begin{aligned}
        &\left(\frac{\omega}{\omega_J}\right)^4+i \frac{\nu_{cn}}{\omega_J}\left(\frac{\omega}{\omega_J}\right)^3+\left[\left(1-\frac{\chi m_d n_{d0}}{4\omega_J^2}k^2-\frac{\omega_{pdc}^2(1-R)}{\omega_J^2(1+k^{-2} \lambda_{D_{r,q}}^{-2})}\right)(k^2+\rho^{-2})k^{-2} \right] \left(\frac{\omega}{\omega_J}\right)^2\\&+i\frac{\nu_{cn}}{\omega_J}\left[\left(1-\frac{\chi m_d n_{d0}}{4\omega_J^2}k^2\right)(k^2+\rho^{-2})k^{-2} \right]\left(\frac{\omega}{\omega_J}\right)-(1-R)\lambda_{D_{r,q}}^2\left(\frac{\omega_{pdc}^2}{\omega_J^2}\right)\\&\times\left(\frac{4 \omega_J^2- 4\omega_{Jc}^2-\chi m_d n_{dn0} k^2}{4\omega_J^2}\right)\left(\frac{k^2+\rho^{-2}}{1+k^2\lambda_{D_{r,q}}^2}\right)k^{-2}=0,
    \end{aligned}
\end{equation}
where, $\omega_{pd}=\sqrt{4\pi q_d^2 n_{d0}/m_d}$ and $\omega_{pdc}=\sqrt{4\pi q_d^2 n_{dc0}/m_{d}}$ are the effective dust-plasma oscillation frequency (contributed by both the neutral and charged dust grains) and the charged dust-plasma oscillation frequency (contributed by the charged dust grains only). The term, $\lambda_{D_{r,q}}$, is the effective dust-plasma Debye length for $(r,q)$-velocity distribution, cast as
\begin{equation}
    \label{eq:3.16}
    \lambda_{D_{r,q}}=\left[\frac{T_e T_i}{4\pi e^2 (An_{e0} T_i+n_{i0} T_e)}\right]^\frac{1}{2},
\end{equation}
which becomes the standard effective plasma Debye length $(\lambda_D)$ for $A=1 (r=0,q\to \infty)$.
\par 
To check the reliability and validation of \eqref{eq:3.15} in comparison with previous predictions found in the literature, all the newly added modifications are turned off, i.e., $R=0,\ \chi=0,\ \rho \to \infty$, and $A=1$. Now, using the approximation $(\lambda_{De}/\lambda_{Di})^2 \gg 1$ \cite{dwivedi1999pulsational}, \eqref{eq:3.15} exactly reduces to the same linear dispersion relation (quartic in degree) in the original form for the PMGC dynamics excited in DMCs obtained by Dwivedi et al. \cite{dwivedi1999pulsational}, as
\begin{equation}
    \label{eq:3.17}
    \begin{aligned}
        \left(\frac{\omega}{\omega_J}\right)^4 +i\frac{\nu_{cn}}{\omega_J}\left(\frac{\omega}{\omega_J}\right)^3+\left(1-\frac{k^2C_\text{scam}^2}{\omega_J^2}\right)&\left(\frac{\omega}{\omega_J}\right)^2+i\frac{\nu_{cn}}{\omega_J}\left(\frac{\omega}{\omega_J}\right)\\&-\frac{k^2C_\text{scam}^2}{\omega_J^2}\left(\frac{1}{1+\eta}\right)=0,
    \end{aligned}
\end{equation}
where, $\eta=n_{dc0}/n_{dn0}$ represents the ratio between the equilibrium charged and neutral dust population densities. $C_\text{scam}=\sqrt{(q_d/e) (n_{dc0} T_i)/(n_{dn0} m_d )}$ refers to the usual phase speed of the supported dust acoustic wave (DAW) or the so-called acoustic mode (SCAM)\cite{dwivedi1999pulsational,dwivedi1997dust}.
\par
To facilitate the analysis of the intricate PMGC fluctuations, a standard astrophysical normalization procedure \cite{kalita2021jeans,dasgupta2019jeans} is employed to \eqref{eq:3.18}, which converts the dispersion relation from previous wave space $(k,\omega)$ to a new wave space $(K,\Omega)$ as follows
\begin{equation}
    \label{eq:3.18}
    \Omega^4+ a_3 \Omega^3+a_2\Omega^2+a_1\Omega+a_0=0;
\end{equation}
\begin{equation}
    \label{eq:3.19}
    a_3 = i \left(\frac{\nu_{cn}}{\omega_J}\right),
\end{equation}
\begin{equation}
    \label{eq:3.20}
    a_2 = \left(1-\frac{\chi m_d n_{d0}}{4 C_{DA}^2}K^2-\frac{(1-R) \omega_J^2 \lambda_{D_{r,q}}^2 \Omega_{pdc}^2}{K^2 \omega_J^2 \lambda_{D_{r,q}}^2+C_{DA}^2 }K^2 \right)\left(K^2+\frac{1}{\xi^2} \right) K^{-2},
\end{equation}
\begin{equation}
    \label{eq:3.21}
    a_1=i \left(\frac{\nu_{cn}}{\omega_J}\right)\left[\left(1-\frac{\chi m_d n_{d0}}{4 C_{DA}^2}K^2 \right)\left(K^2+\frac{1}{\xi^2} \right) K^{-2} \right],
\end{equation}
\begin{equation}
    \label{eq:3.22}
    \begin{aligned}
    a_0= -(1-R) \Omega_{pdc}^2 \left(\frac{\lambda_{D_{r,q}}}{\lambda_J}\right)^2 \frac{C_{DA}^2}{(K^2 \omega_J^2 \lambda_{D_{r,q}}^2+C_{DA}^2)}& \left(1-\Omega_{Jc}^2-\frac{\chi K^2}{4 C_{DA}^2}m_d n_{dn0} \right)\\& \times \left(K^2+\frac{1}{\xi^2}\right)^2 K^{-2}.
    \end{aligned}
\end{equation}
Here, $\xi=\rho/\lambda_J$ and $K=k (C_{DA}/\omega_J)$ are, respectively, the dimensionless radial distance and angular wavenumber normalized with the Jeans length $(\lambda_J)$ and angular Jeans wavenumber ($k_J\equiv \omega_J/C_{DA}$, $C_{DA}$ being the dust acoustic phase velocity). Similarly, $\Omega = \omega/\omega_J$, $\Omega_{pdc}=\omega_{pdc}/\omega_J$, and $\Omega_{Jc}=\omega_{Jc}/\omega_J$ are, respectively, the dimensionless angular frequency, charged dust-plasma oscillation frequency, and critical Jeans frequency for charged dust normalized with the Jeans angular frequency $(\omega_J)$.
\par 
Within the realm of ultra low-frequency (ULF) fluctuations, \eqref{eq:3.19} transforms to an equation characterized by a vanishing propagatory component $(\Omega_r =0)$, signifying the wave condensation and a non-vanishing decay (growth) component $(\Omega_i \neq 0)$, indicating wave collapse \cite{karmakar2017evolutionary}. This implies that the wave no longer exhibits propagatory characteristics in the sense of classical wave mechanics. However, the wave undergoes either exponential decay or growth over time, even in the absence of propagation. Hence, the final dispersion relation after separating the real $(\Omega_r)$ and imaginary $(\Omega_i)$ parts can be written as
\begin{equation}
    \label{eq:3.23}
    \Omega_i = \frac{(1-R)(1+K^2\xi^2)(-4C_{DA}^2\omega_J^2+ \chi m_d n_{dn0}\omega_J^2 K^2 +4C_{DA}^2\omega_{Jc}^2)C_{DA}^2\lambda_{D_{r,q}}^2\Omega_{pdc}^2}{\nu_{cn}\omega_J\lambda_J^2\xi^2(4C_{DA}^2- \chi m_d n_{d0} K^2)(C_{DA}^2+K^2\lambda_{D_{r,q}}^2\omega_J^2)}.
\end{equation}
Equation \eqref{eq:3.23} is the required linear dispersion relation for the PMGC instability in a self-gravitating complex DMC due to the combined influence of the polarization force and $(r,q)$-distributed electrons within the EiBI gravity framework.

\section{Results and discussions}\label{sec:Results and Discussions}
\begin{figure}[htbp]
    \centering
    \includegraphics[width=.6\textwidth]{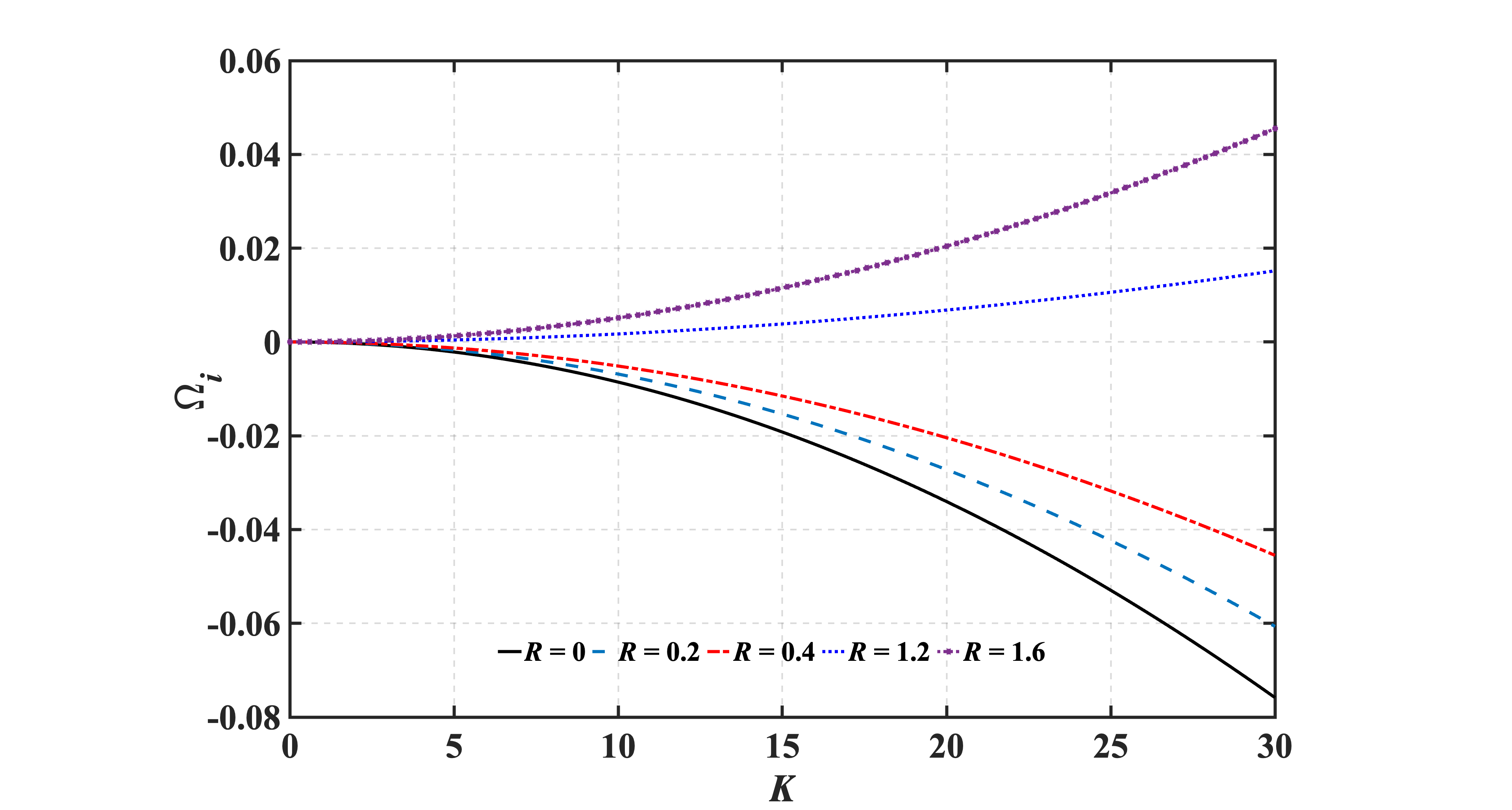}
    \caption{Variation of the growth rate $(\Omega_i)$ with the angular wavenumber $(K)$ for different values of the polarization parameter $(R)$ on the Jeans scale. The various curves link to: (a) $R = 0$ (black solid line), (b) $R = 0.2$ (cyan dashed line), (c) $R = 0.4$ (red dashed-dotted line), (d) $R = 1.2$ (blue dotted line), and $R = 1.6$ (magenta dotted pentagram line), respectively.}
    \label{fig:figure2}
\end{figure}

\begin{figure}[htbp]
    \centering 
    \begin{minipage}{0.49\textwidth}
        \centering
        \includegraphics[trim={4.3cm} {0 cm} {4.3cm} {0cm},clip, width=1\linewidth]{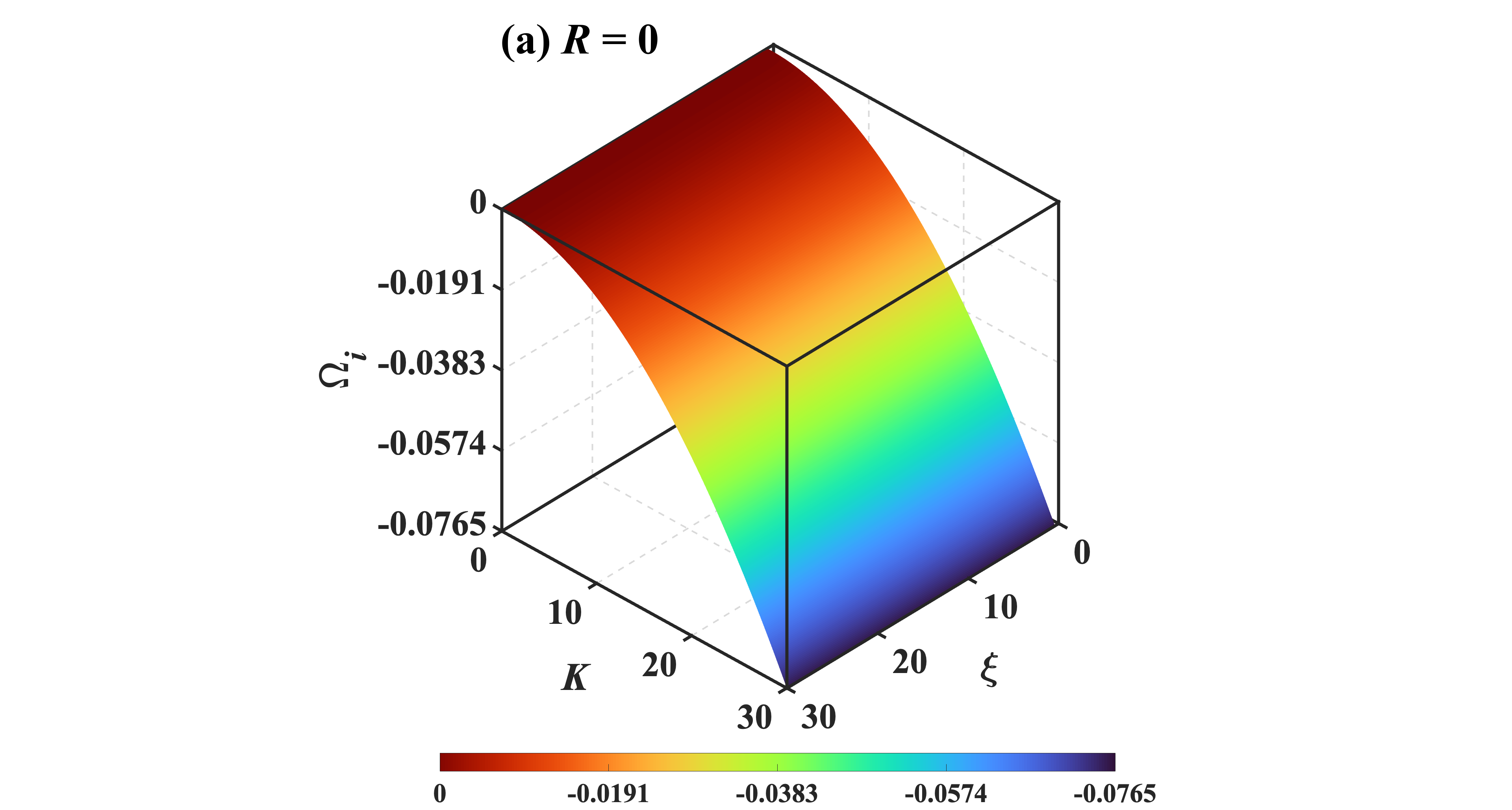}
    \end{minipage}
    \hfill
    \vspace{0.5cm}
    \begin{minipage}{0.49\textwidth}
        \centering
        \includegraphics[trim={4.3cm} {0 cm} {4.3cm} {0cm},clip, width=1\linewidth]{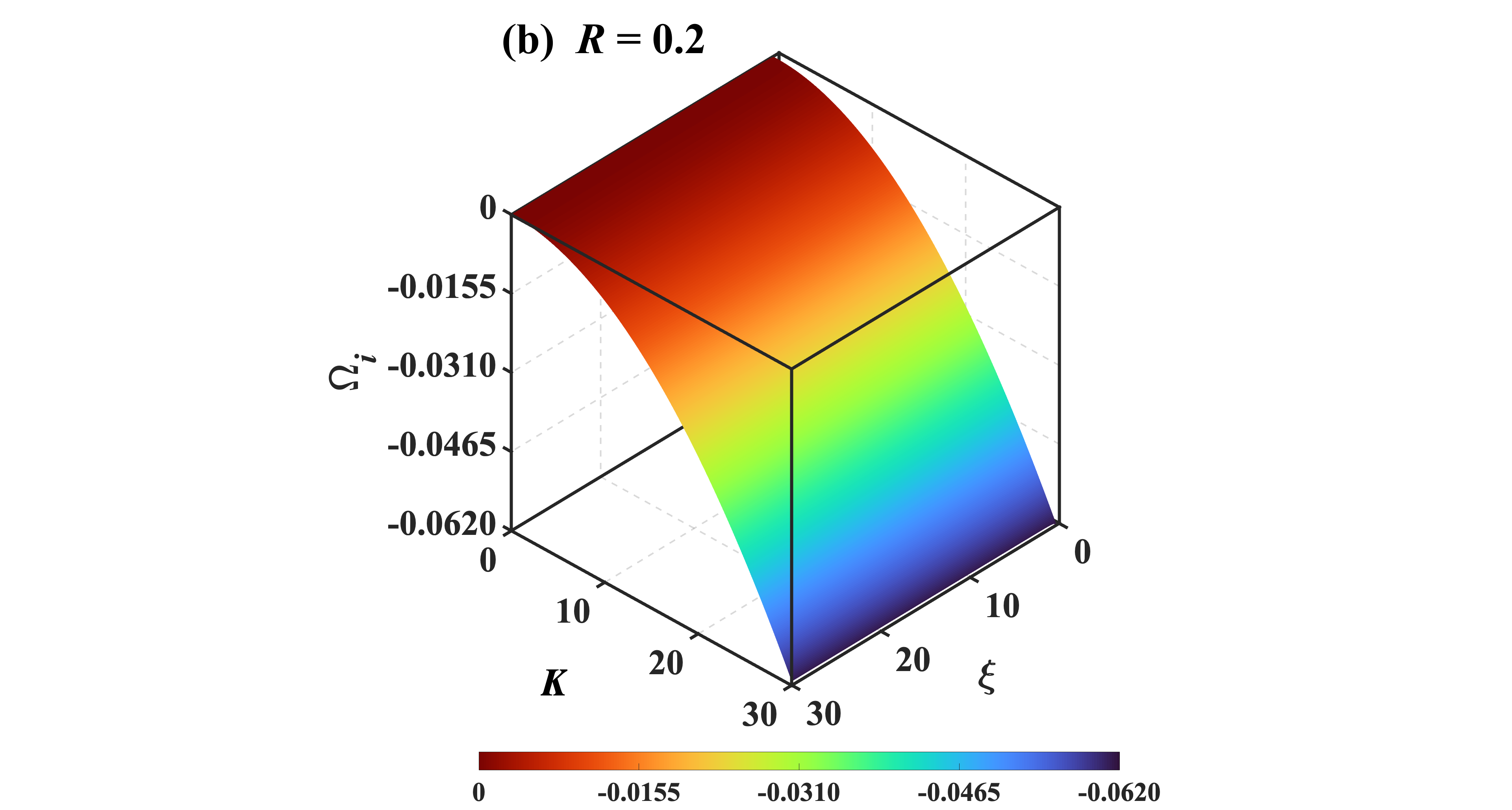}
    \end{minipage}
    \begin{minipage}{0.49\textwidth}
        \centering
        \includegraphics[trim={4.3cm} {0 cm} {4.3cm} {0cm},clip, width=1\linewidth]{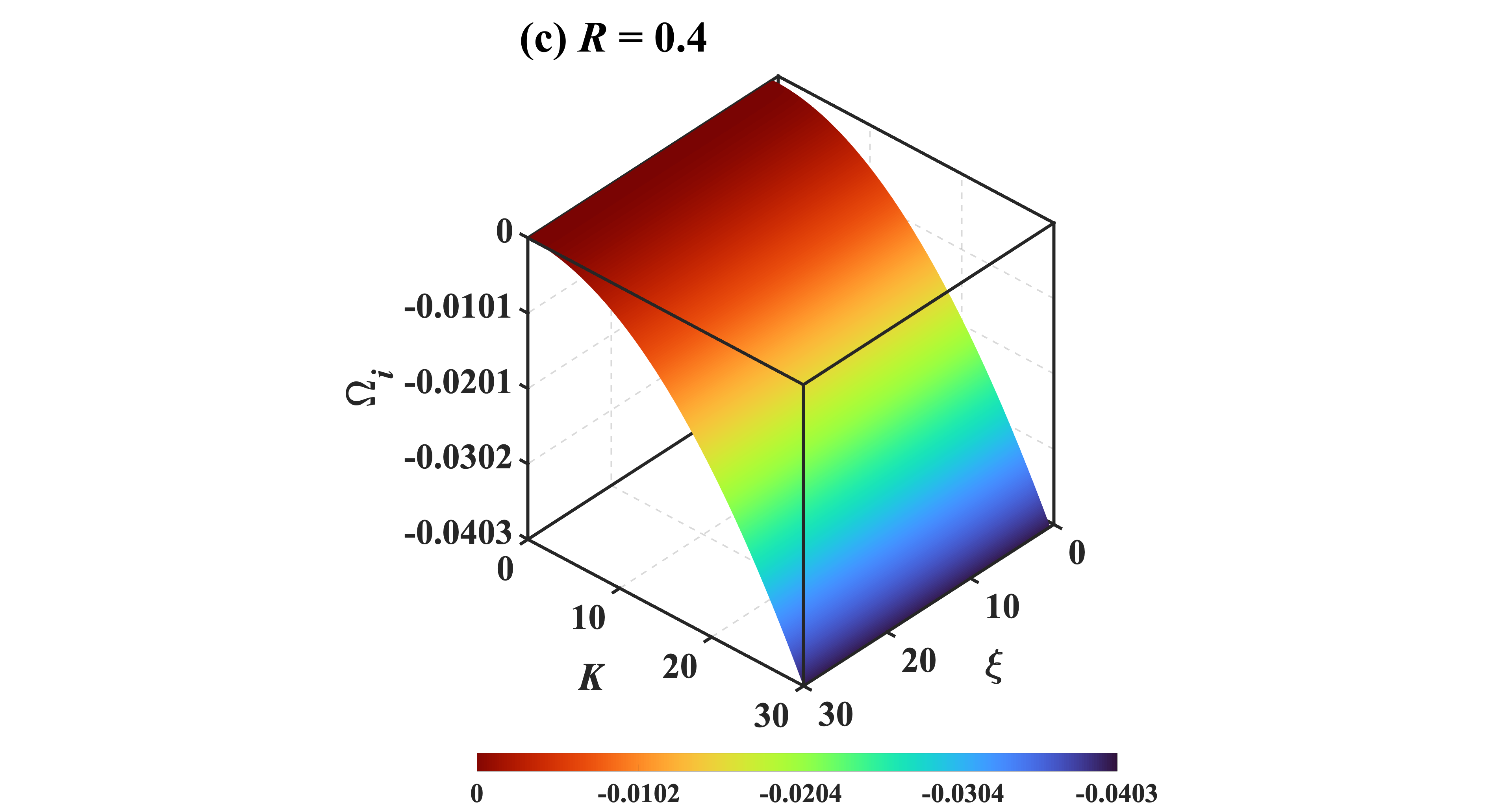}
    \end{minipage}
    \hfill
    \begin{minipage}{0.49\textwidth}
        \centering
        \includegraphics[trim={4.3cm} {0 cm} {4.3cm} {0cm},clip, width=1\linewidth]{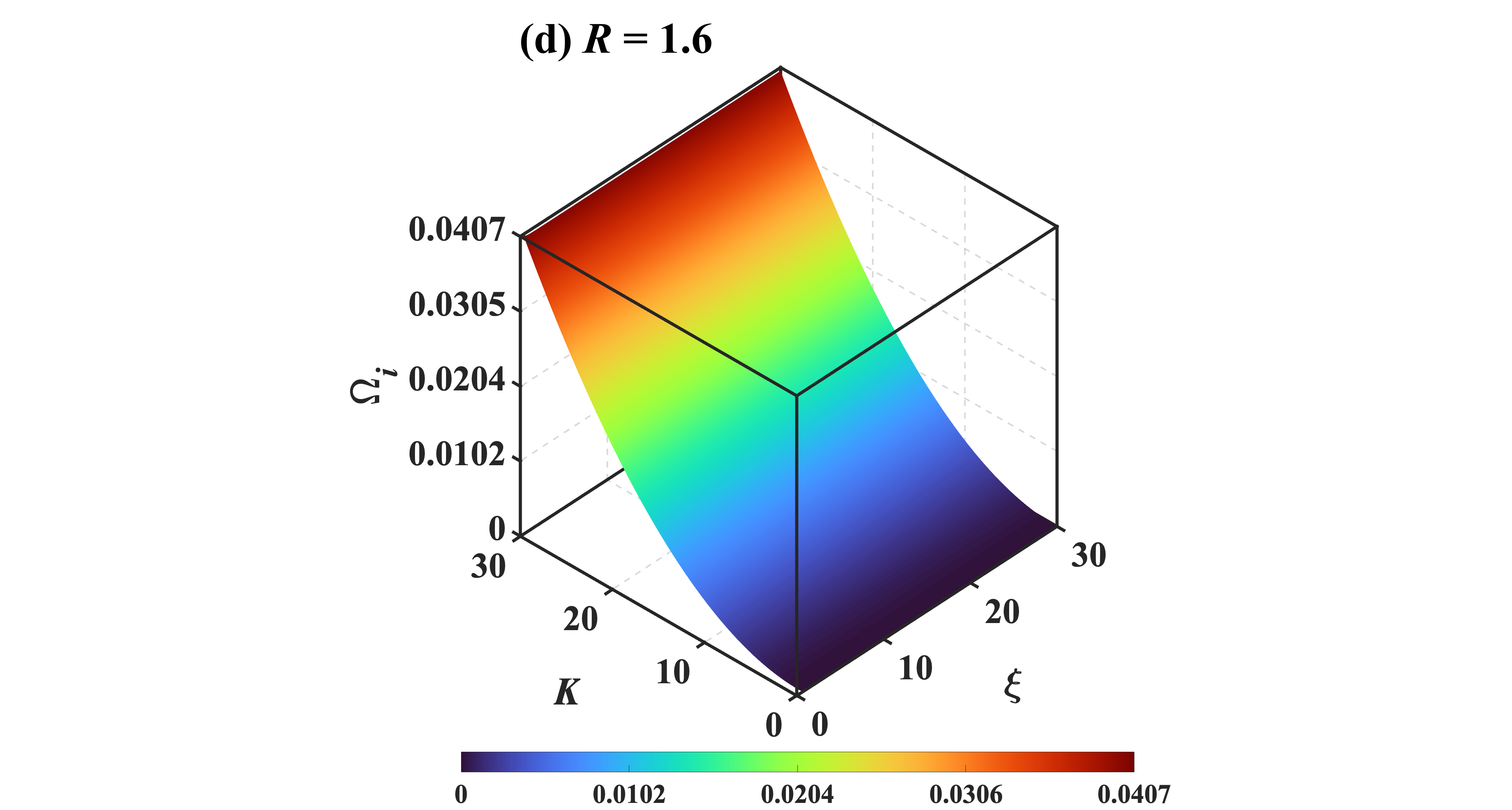}
    \end{minipage}
    \caption{Same as figure \ref{fig:figure2}, but in a wave space defined by Jeans-scaled radial distance $(\xi)$ and Jeans-scaled angular wavenumber $(K)$. The different panels correspond to (a) $R = 0$, (b) $R = 0.2$, (c) $R = 0.4$, and (d) $R = 1.6$, respectively.}
    \label{fig:figure3}
\end{figure}

\begin{figure}[htbp]
    \centering 
    \begin{minipage}{0.49\textwidth}
        \centering
        \includegraphics[trim={4.3cm} {0 cm} {4.2cm} {0cm},clip,width=1\linewidth]{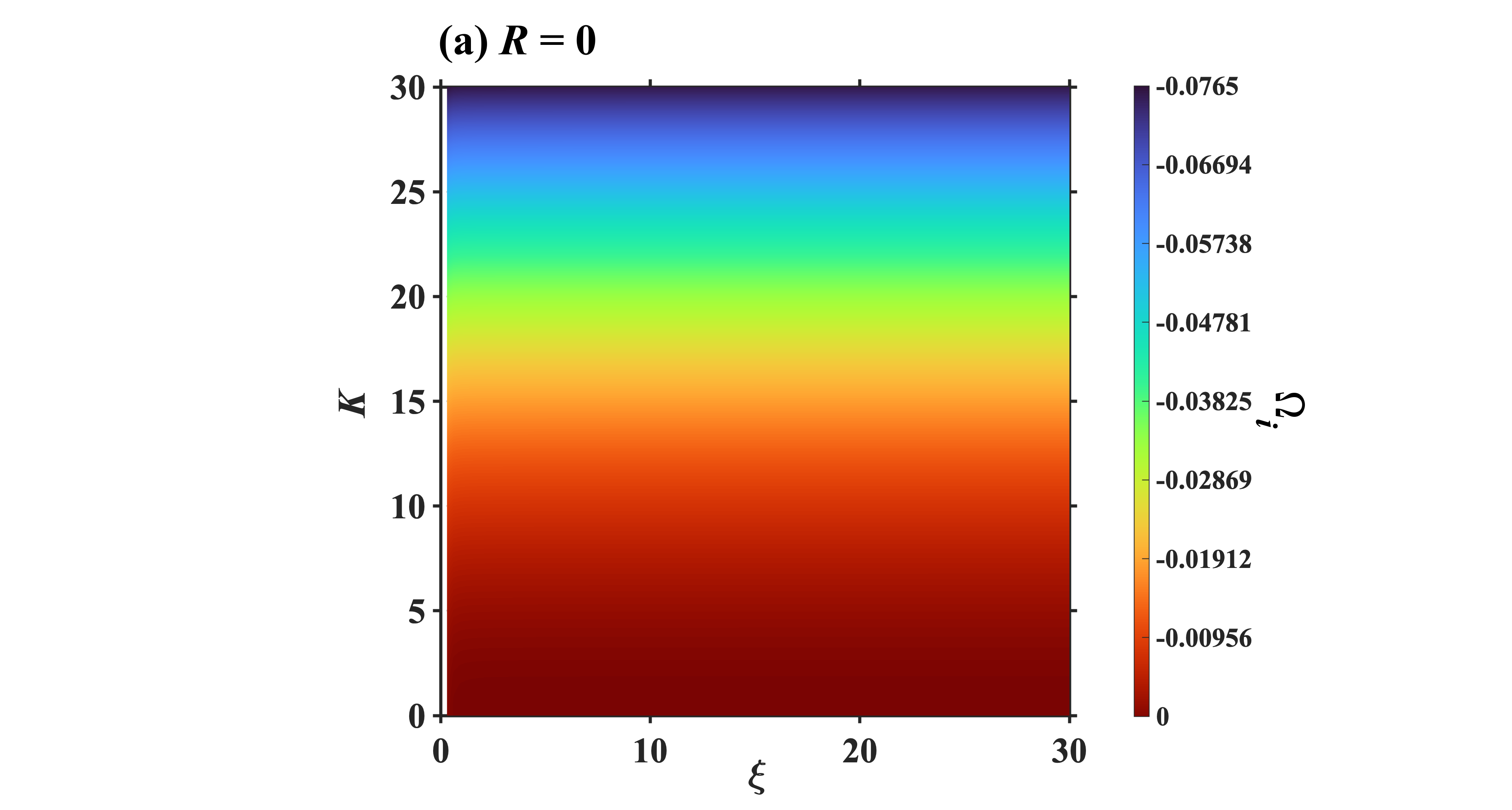}
    \end{minipage}
    \hfill
    \vspace{0.5cm}
    \begin{minipage}{0.49\textwidth}
        \centering
        \includegraphics[trim={4.3cm} {0 cm} {4.3cm} {0cm},clip, width=1\linewidth]{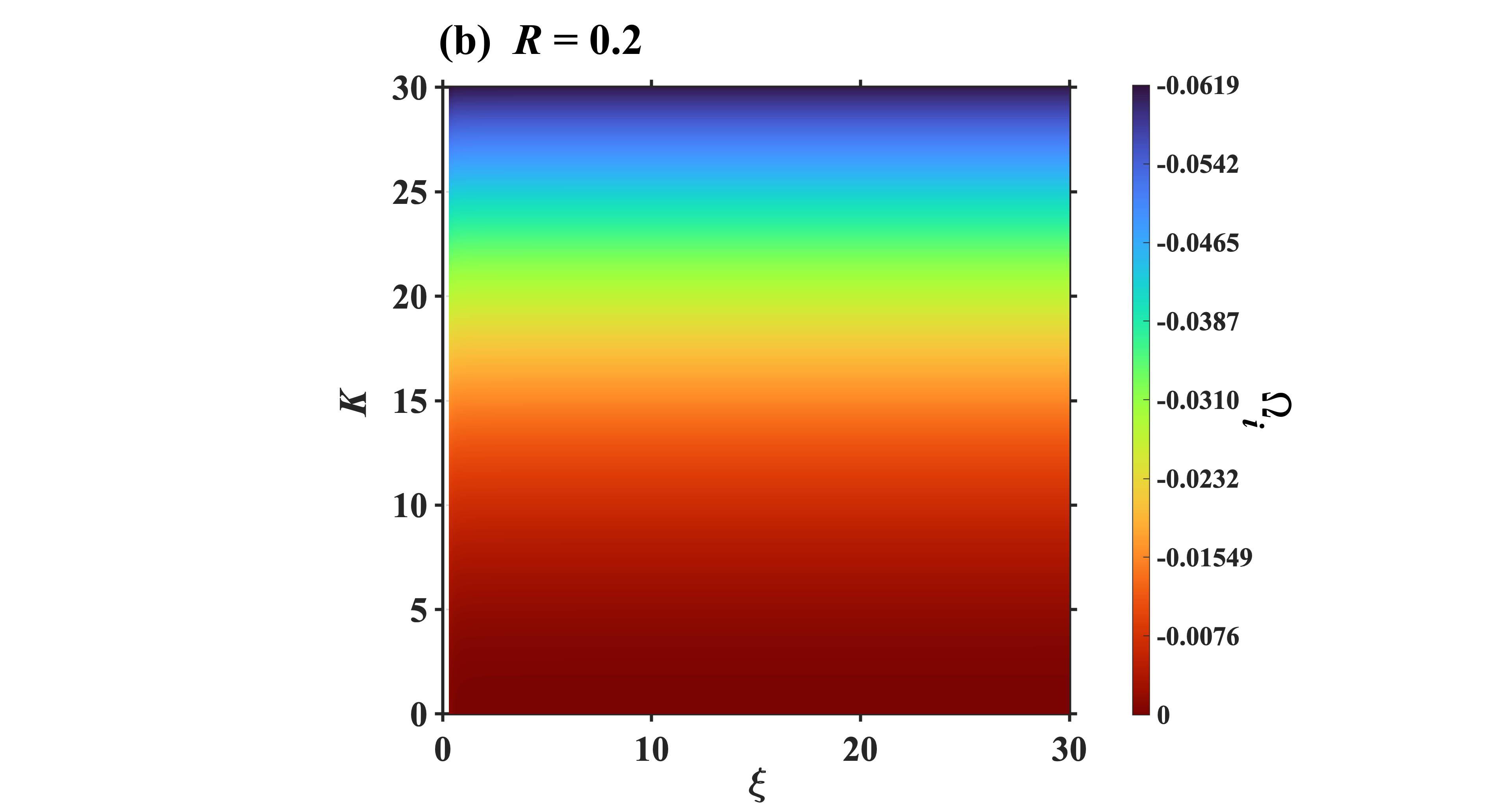}
    \end{minipage}
    \begin{minipage}{0.49\textwidth}
        \centering
        \includegraphics[trim={4.3cm} {0 cm} {4.3cm} {0cm},clip, width=1\linewidth]{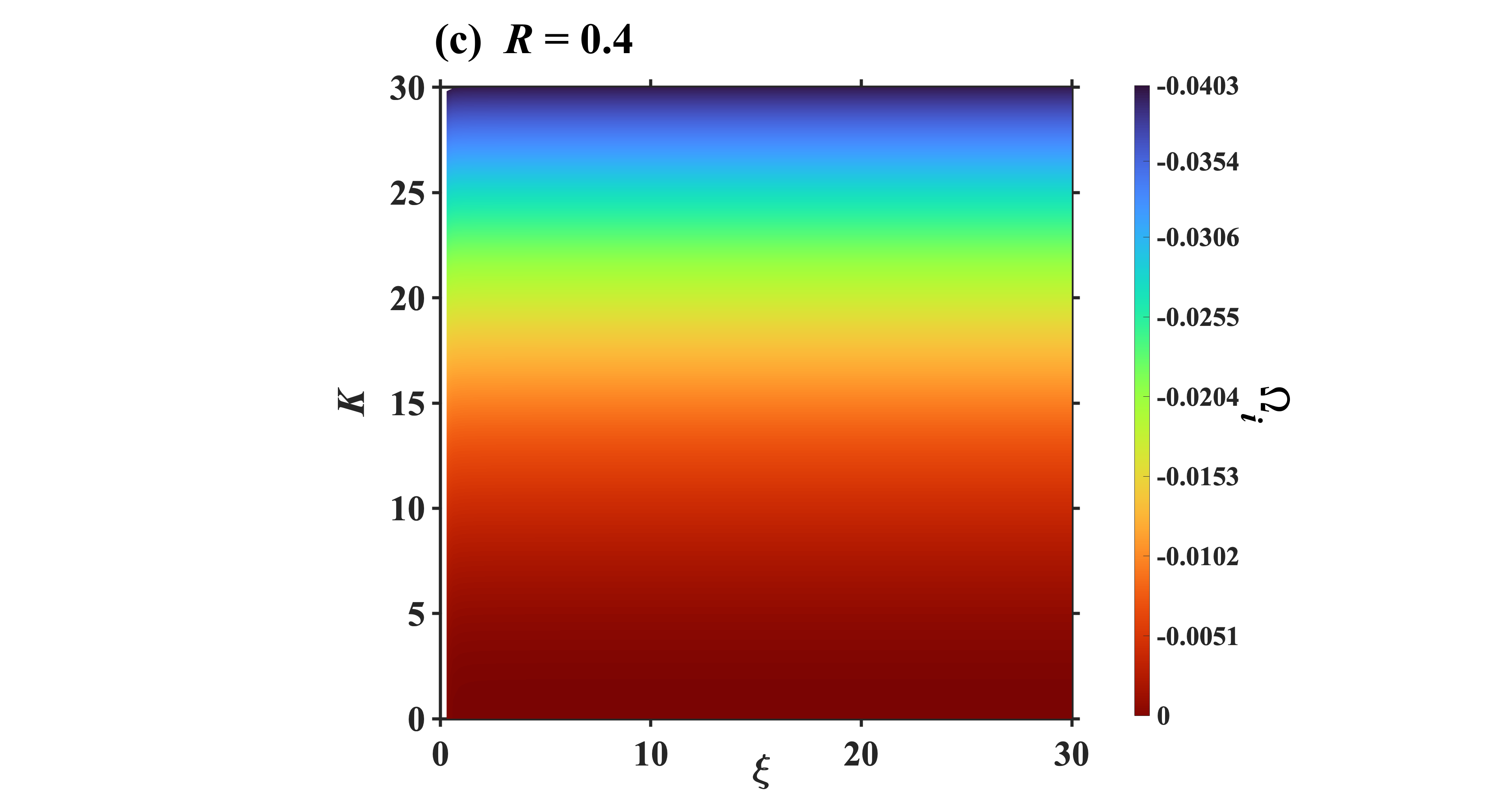}
    \end{minipage}
    \hfill
    \begin{minipage}{0.49\textwidth}
        \centering
        \includegraphics[trim={4.3cm} {0 cm} {4.2cm} {0cm},clip, width=1\linewidth]{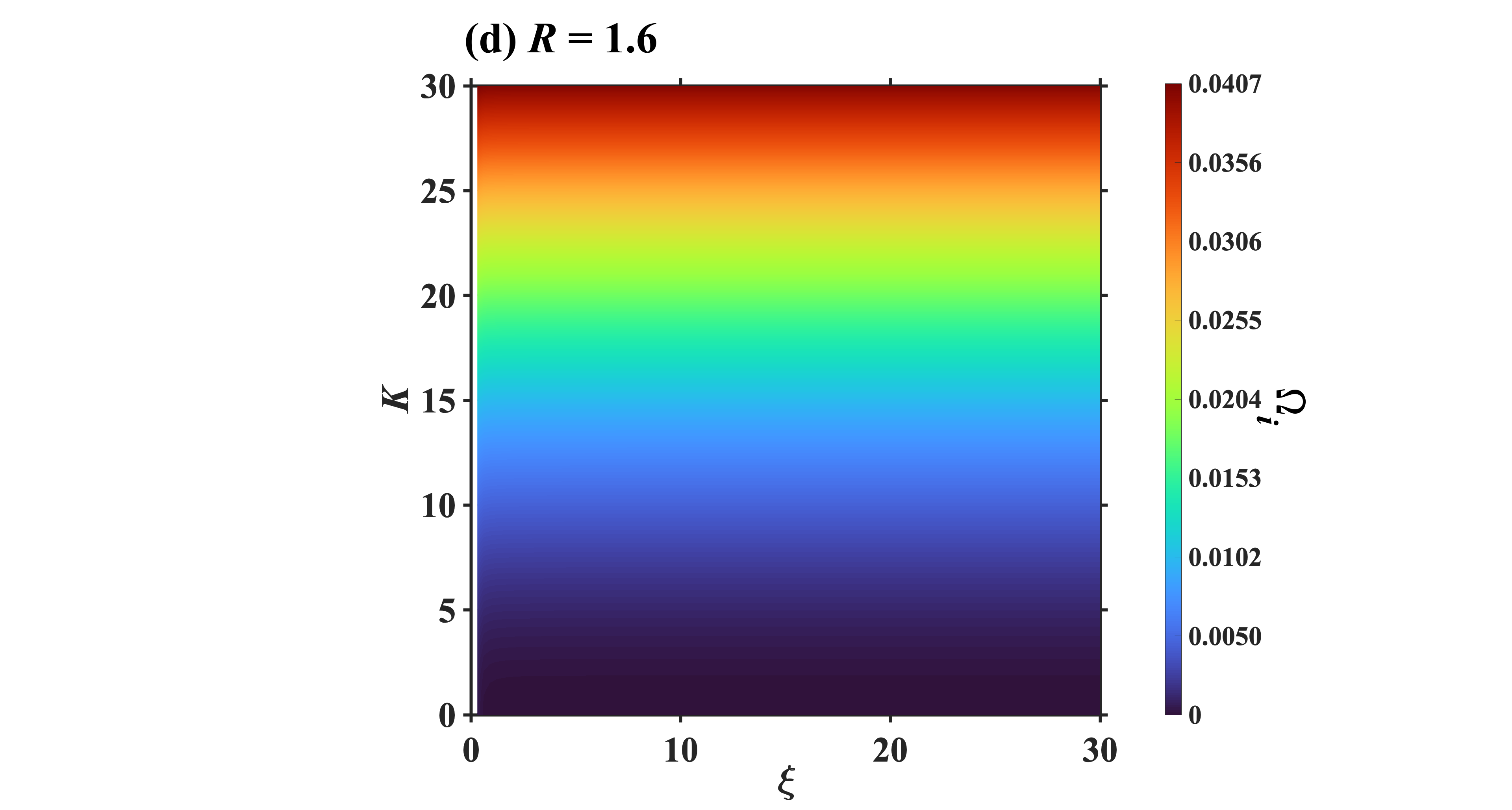}
    \end{minipage}
    \caption{Same as figure \ref{fig:figure3}, but in a color-spectral top-view projection.}
    \label{fig:figure4}
\end{figure}

\begin{figure}[htbp]
    \centering 
    \begin{minipage}{0.49\textwidth}
        \centering
        \includegraphics[trim={4.3cm} {0 cm} {4.3cm} {0cm},clip, width=1\linewidth]{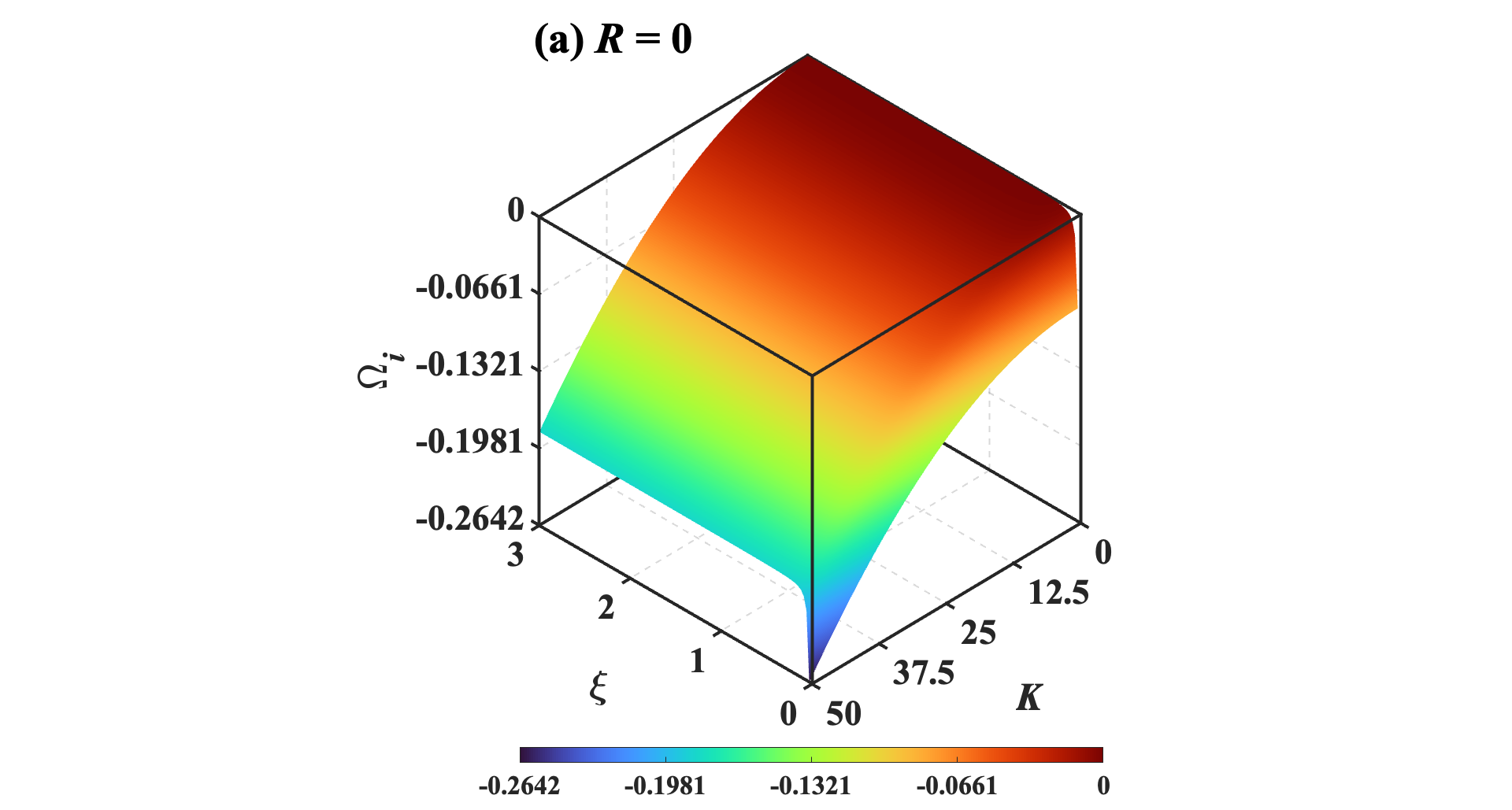}
    \end{minipage}
    \hfill
    \vspace{0.5cm}
    \begin{minipage}{0.49\textwidth}
        \centering
        \includegraphics[trim={4.3cm} {0 cm} {4.3cm} {0cm},clip, width=1\linewidth]{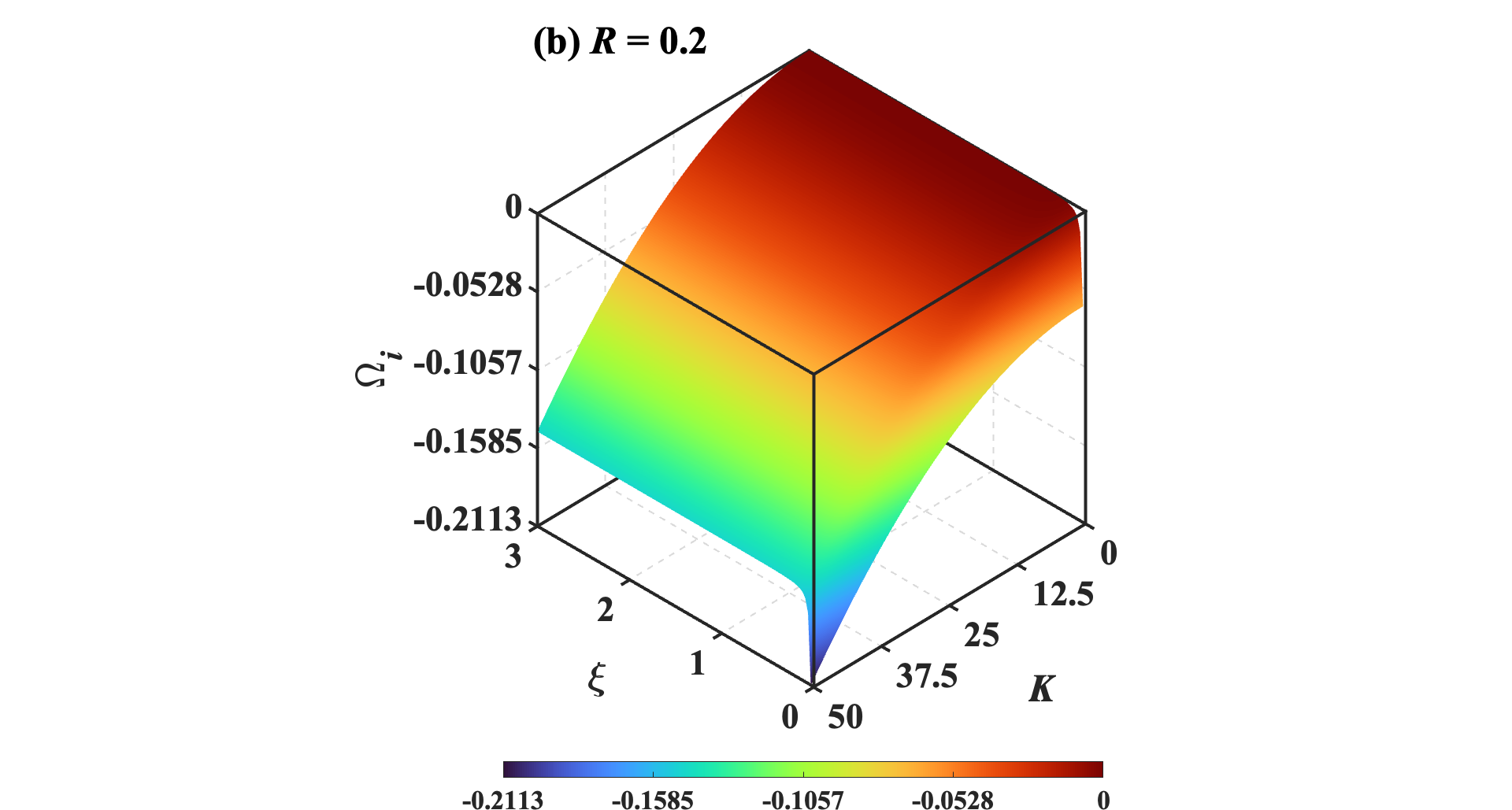}
    \end{minipage}

    \begin{minipage}{0.49\textwidth}
        \centering
        \includegraphics[trim={4.3cm} {0 cm} {4.3cm} {0cm},clip, width=1\linewidth]{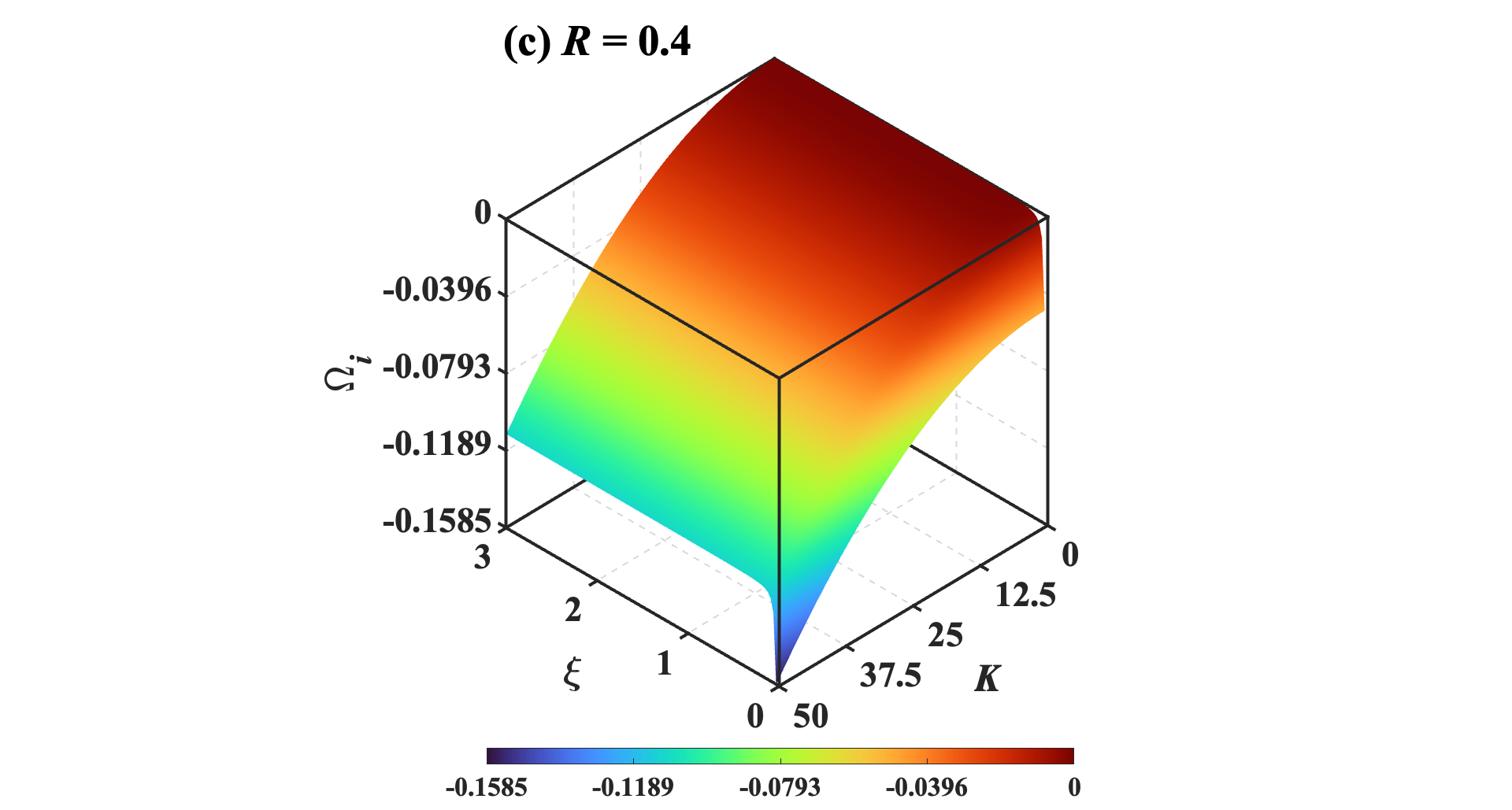}
    \end{minipage}
    \hfill
    \begin{minipage}{0.49\textwidth}
        \centering
        \includegraphics[trim={4.3cm} {0 cm} {4.3cm} {0cm},clip, width=1\linewidth]{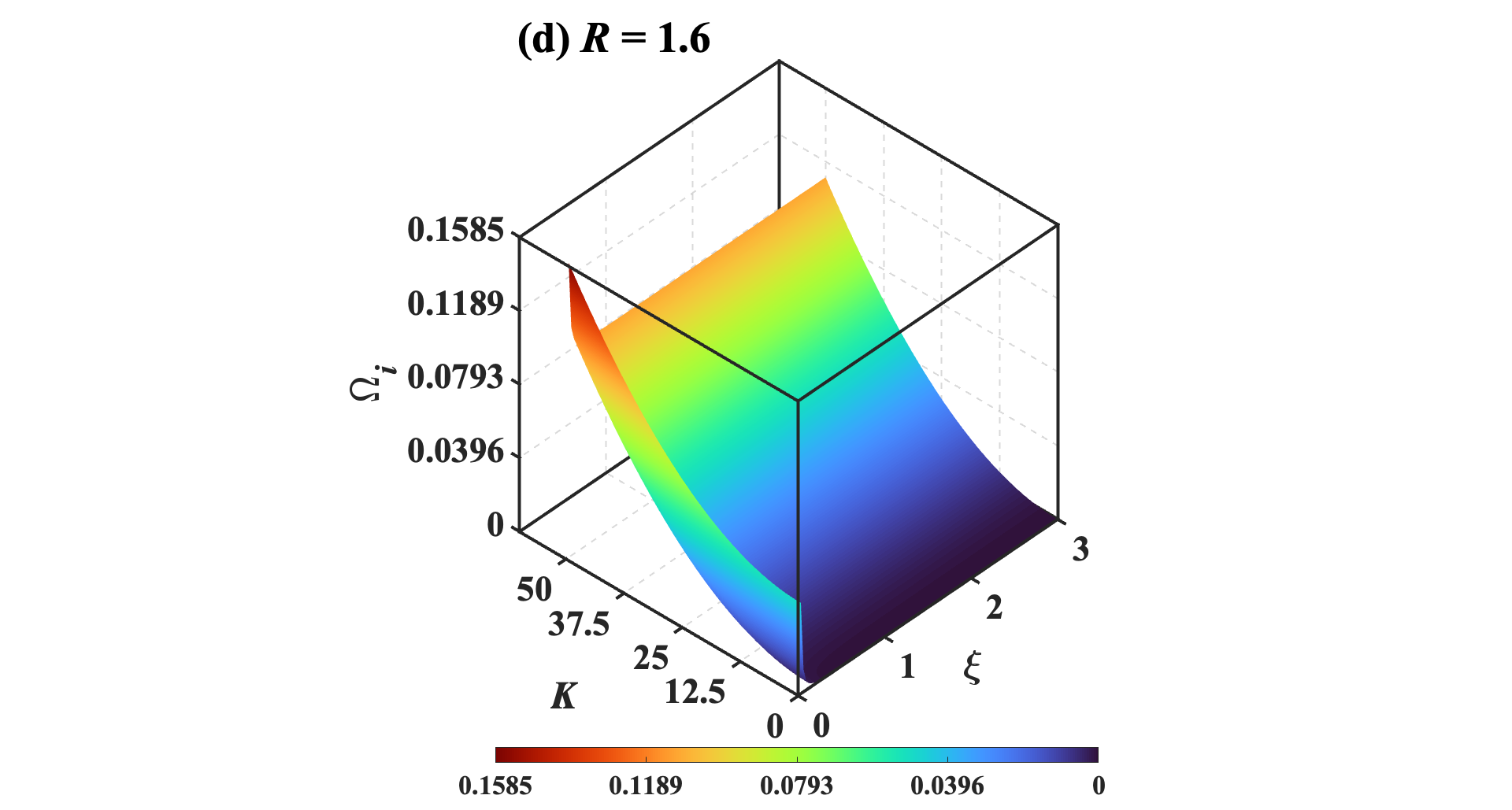}
    \end{minipage}
    \caption{Same as figure \ref{fig:figure3}, but in a zoomed-in form $(\xi = 0-3$, and $K = 0-50)$.}
    \label{fig:figure5}
\end{figure}

\begin{figure}[htbp]
    \centering
    \includegraphics[width=.6\textwidth]{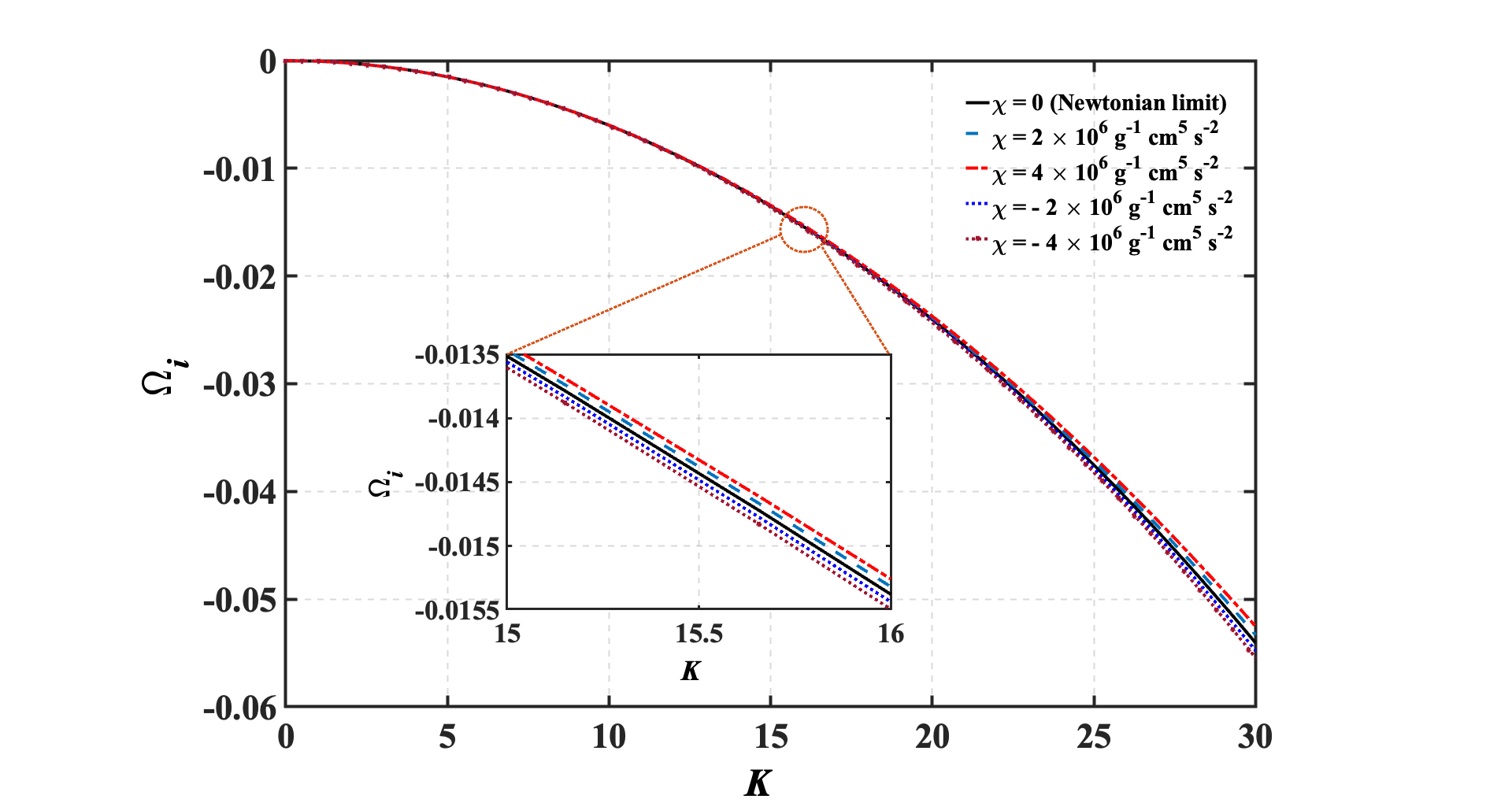}
    \caption{Same as figure \ref{fig:figure2}, but for different values of the EiBI parameter $(\chi)$. The various curves respectively link to: (a) $\chi = 0$ (black solid line, taken as the Newtonian reference line), (b) $\chi=2\times10^6$ \unit{g^{-1}.cm^5.s^{-2}} (cyan dashed line), (c) $\chi=4 \times 10^6$ \unit{g^{-1}.cm^5.s^{-2}} (red dashed-dotted line), (d) $\chi=-2 \times10^6$ \unit{g^{-1}.cm^5.s^{-2}} (blue dotted line), and (e) $\chi= -4\times10^6$ \unit{g^{-1}.cm^5.s^{-2}} (magenta dotted pentagram line).}
    \label{fig:figure6}
\end{figure}

\begin{figure}[htbp]
    \centering 
    \begin{minipage}{0.49\textwidth}
        \centering
        \includegraphics[trim={4.3cm} {0 cm} {4.3cm} {0cm},clip, width=1\linewidth]{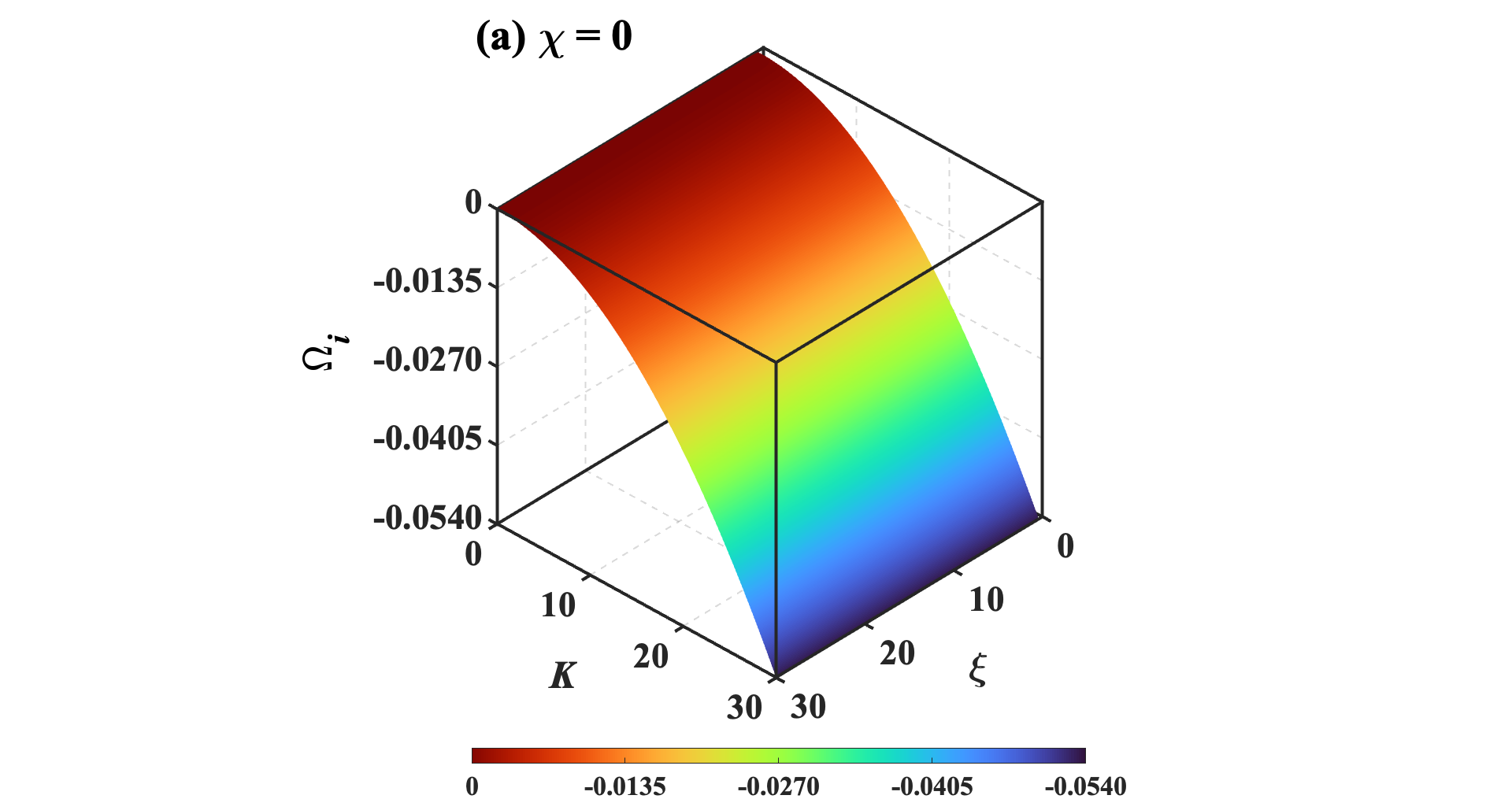}
    \end{minipage}
    \hfill
    \vspace{0.5cm}
    \begin{minipage}{0.49\textwidth}
        \centering
        \includegraphics[trim={4.3cm} {0 cm} {4.3cm} {0cm},clip, width=1\linewidth]{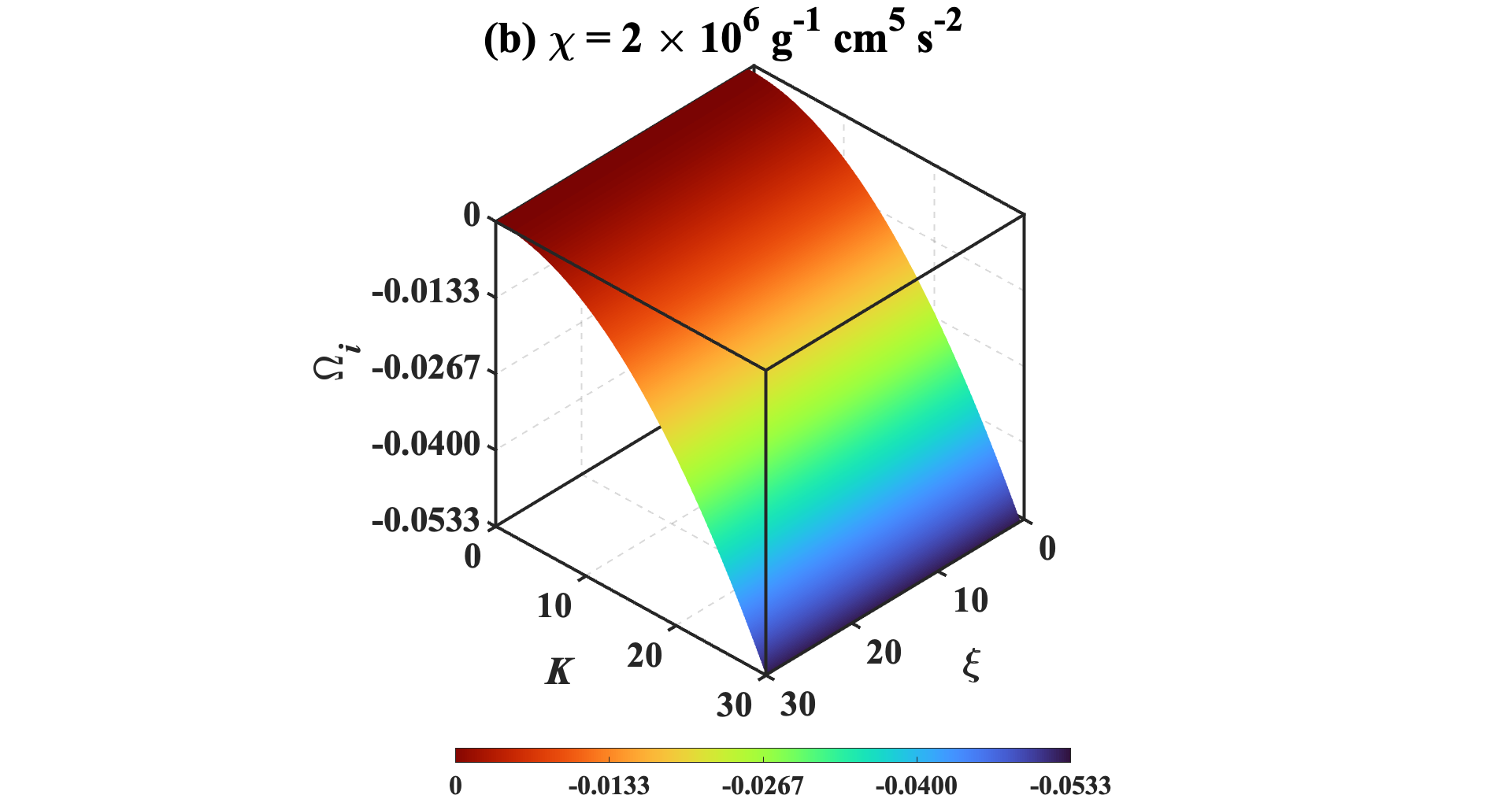}
    \end{minipage}
    \begin{minipage}{0.5\textwidth}
        \centering
        \includegraphics[trim={4.3cm} {0 cm} {4.3cm} {0cm},clip, width=1\linewidth]{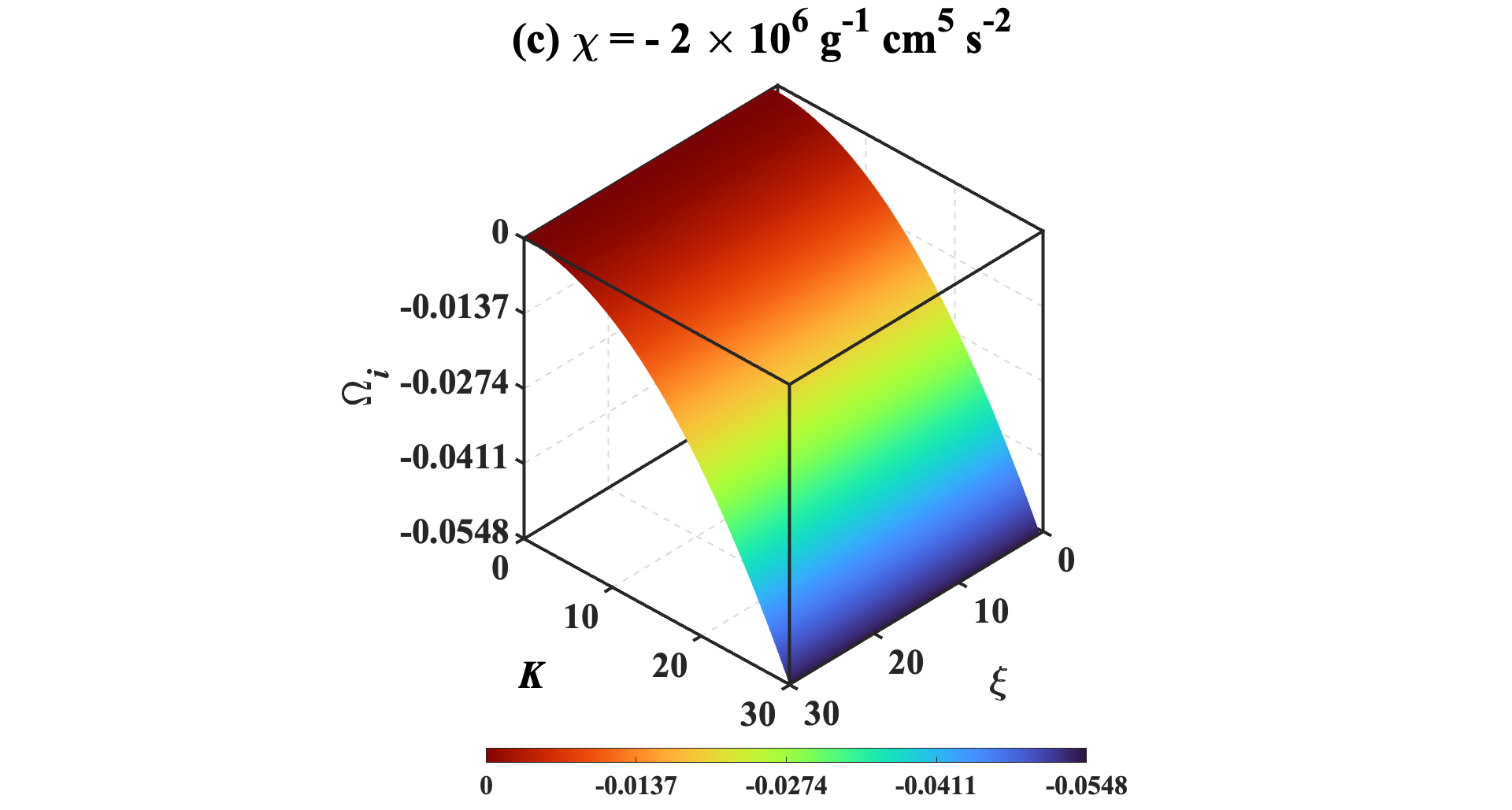}
    \end{minipage}
    \caption{Same as figure \ref{fig:figure6}, but in a wave space defined by Jeans-scaled radial distance $(\xi)$ and Jeans-scaled angular wavenumber $(K)$. The different panels correspond to (a) $\chi = 0$ (taken as the Newtonian reference), (b) $\chi= 2\times10^6$ \unit{g^{-1}.cm^5.s^{-2}}, and (c) $\chi= -2\times10^6$ \unit{g^{-1}.cm^5.s^{-2}}, respectively.}
    \label{fig:figure7}
\end{figure}

\begin{figure}[htbp]
    \centering 
    \begin{minipage}{0.49\textwidth}
        \centering
        \includegraphics[trim={4.3cm} {0 cm} {4.3cm} {0cm},clip, width=1\linewidth]{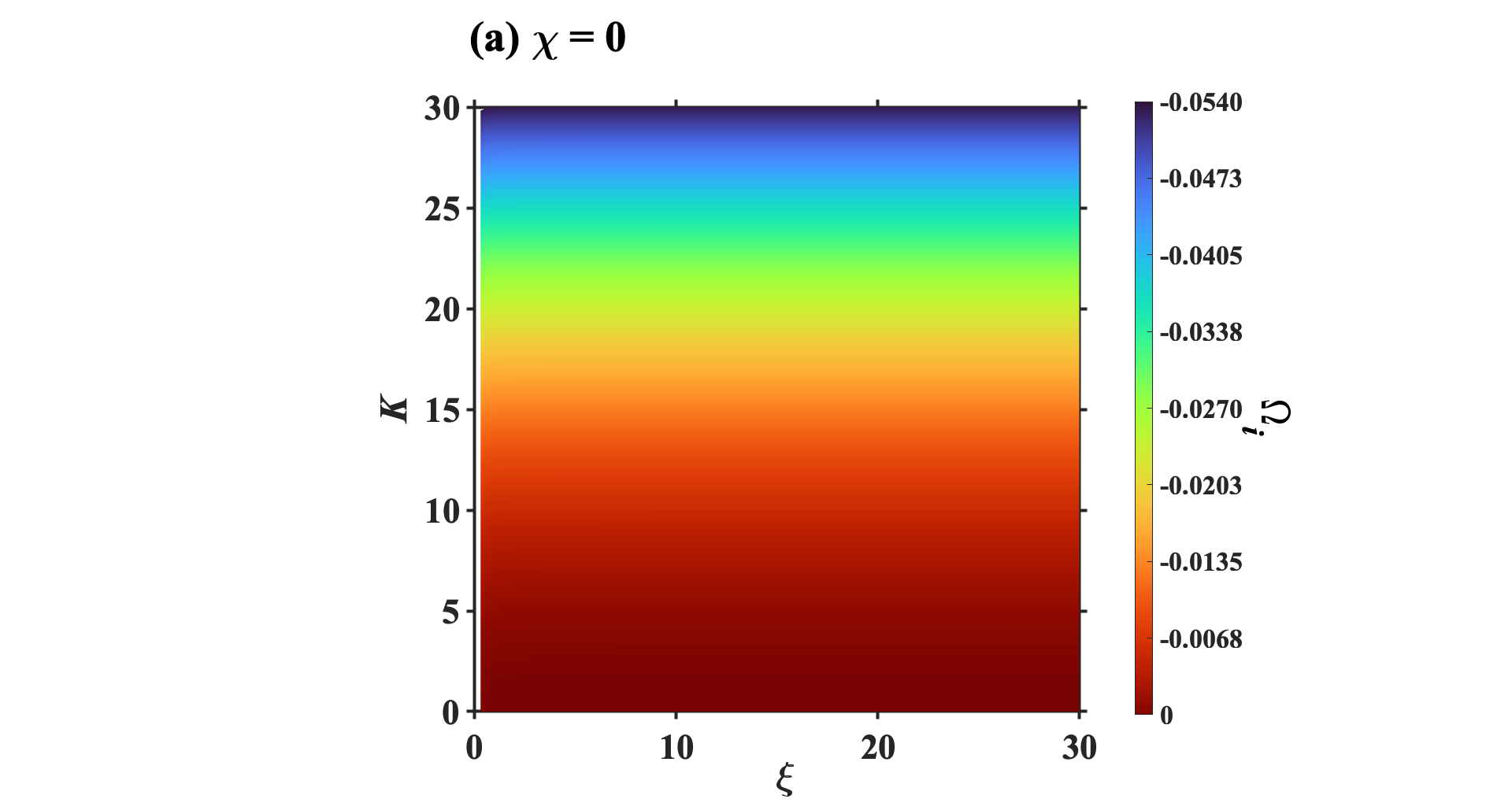}
    \end{minipage}
    \hfill
    \vspace{0.5cm}
    \begin{minipage}{0.49\textwidth}
        \centering
        \includegraphics[trim={4.3cm} {0 cm} {4.3cm} {0cm},clip, width=1\linewidth]{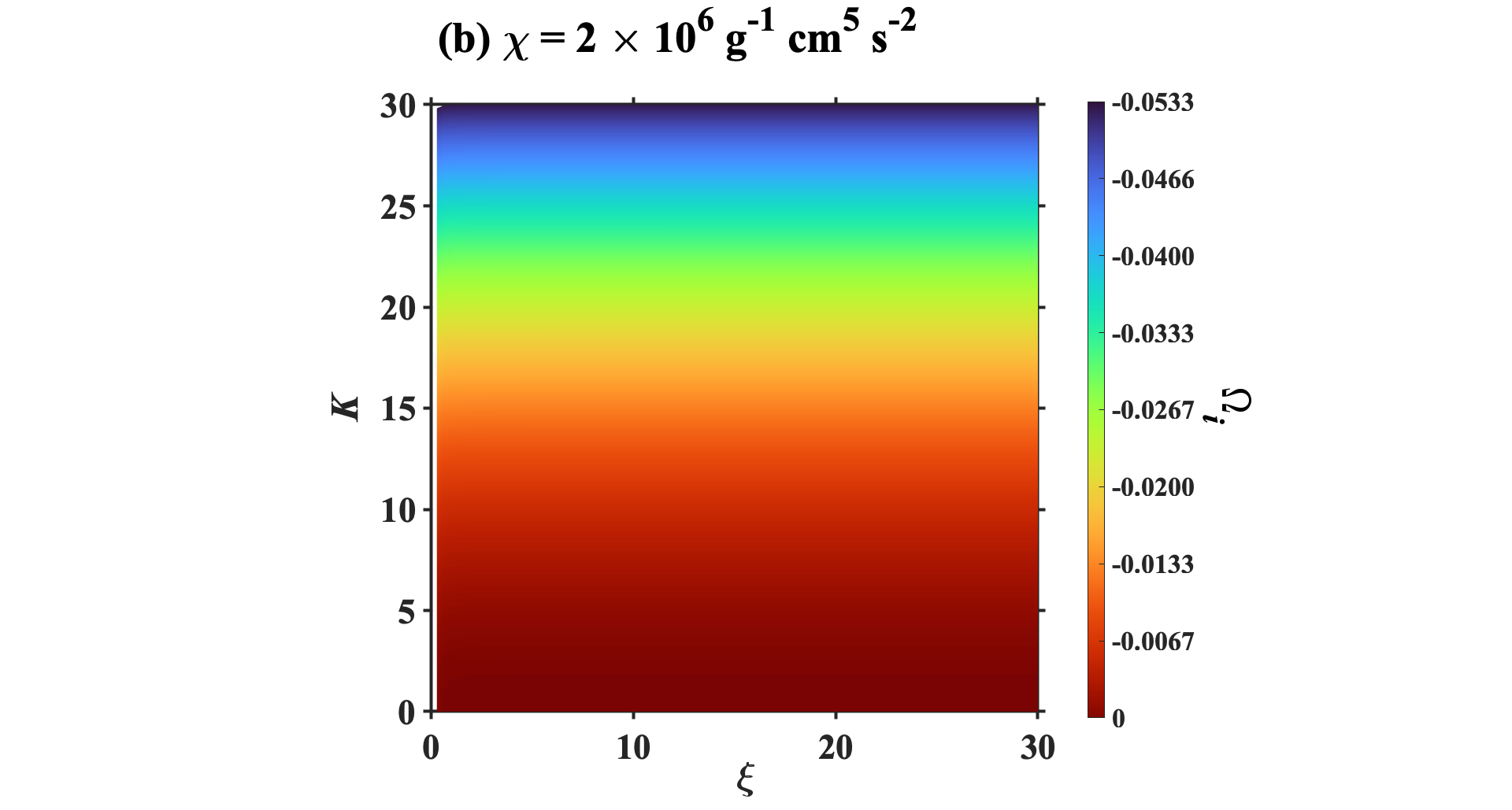}
    \end{minipage}
    \begin{minipage}{0.5\textwidth}
        \centering
        \includegraphics[trim={4.3cm} {0 cm} {4.3cm} {0cm},clip, width=1\linewidth]{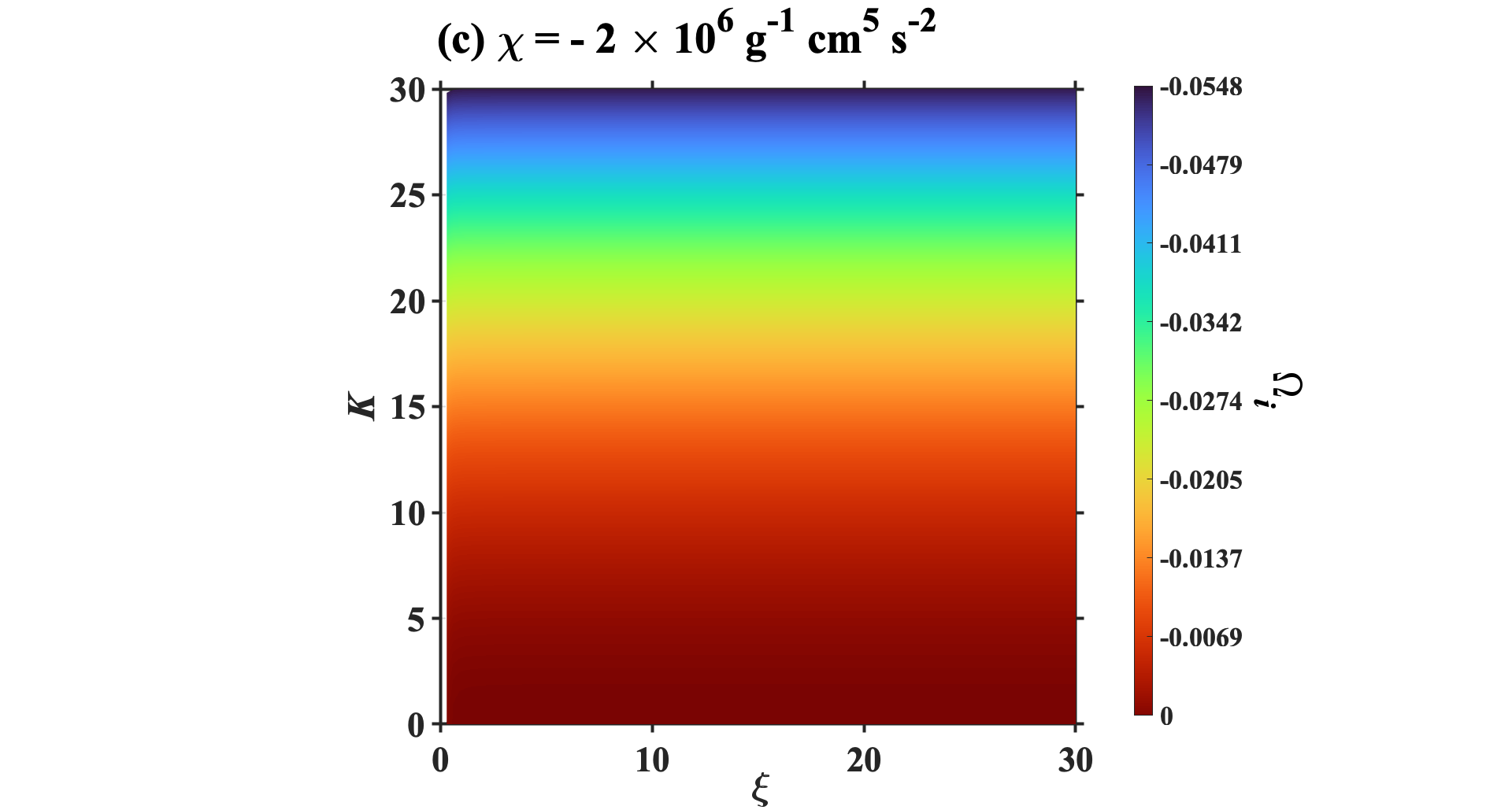}
    \end{minipage}
    \caption{Same as figure \ref{fig:figure7}, but in a colour-spectral top-view projection.}
    \label{fig:figure8}
\end{figure}

\begin{figure}[htbp]
    \centering
    \begin{minipage}{0.49\textwidth}
        \centering
        \includegraphics[width=1\textwidth]{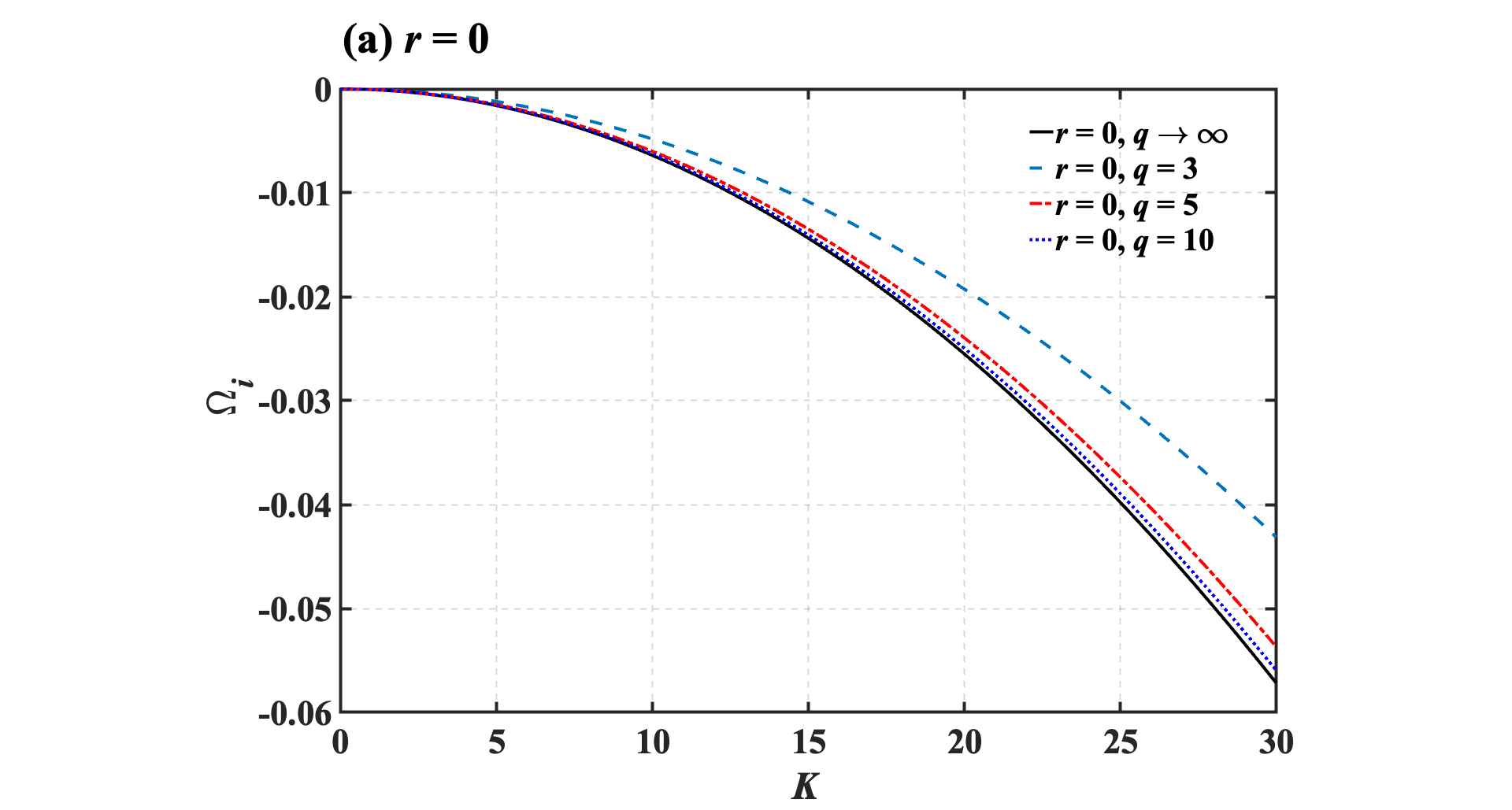}
    \end{minipage}
    \hfill
    \begin{minipage}{0.49\textwidth}
        \centering
        \includegraphics[width=1\textwidth]{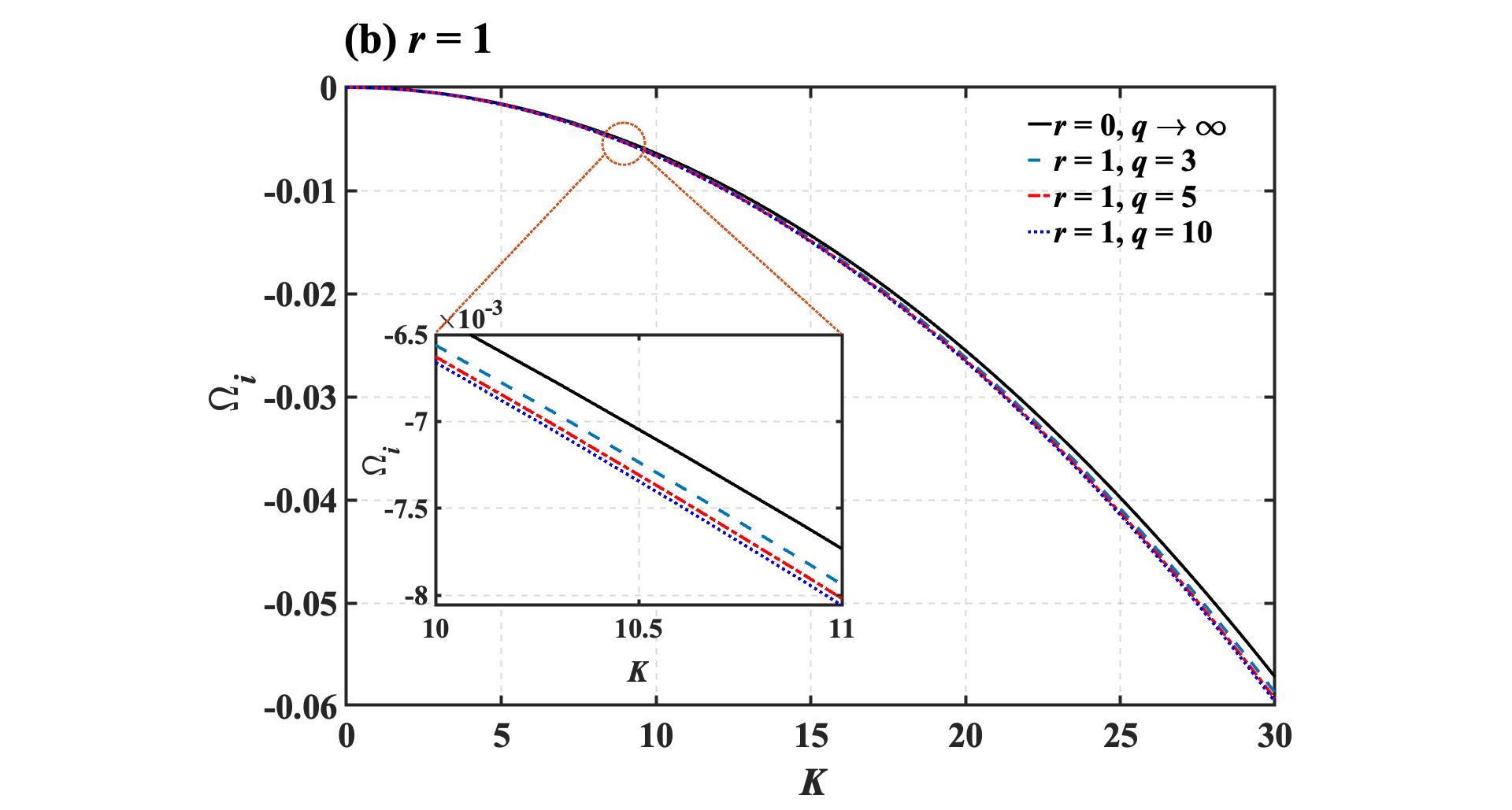}
    \end{minipage}
    \caption{Same as figure \ref{fig:figure2}, but for different values of $q$ with (a) $r = 0$ and (b) $r=1$. The various curves respectively link to : (a) $r = 0,\ q \to \infty $ (black solid line, taken as the Maxwellian reference line), (b) $q = 3$ (cyan dashed line), (c) $q = 5$ (red dashed-dotted line), and (d) $q = 10$ (blue dotted line).}
    \label{fig:figure9}
\end{figure}

\begin{figure}[htbp]
    \centering
    \begin{subfigure}{0.49\textwidth}
    \centering
    \includegraphics[width=1\textwidth]{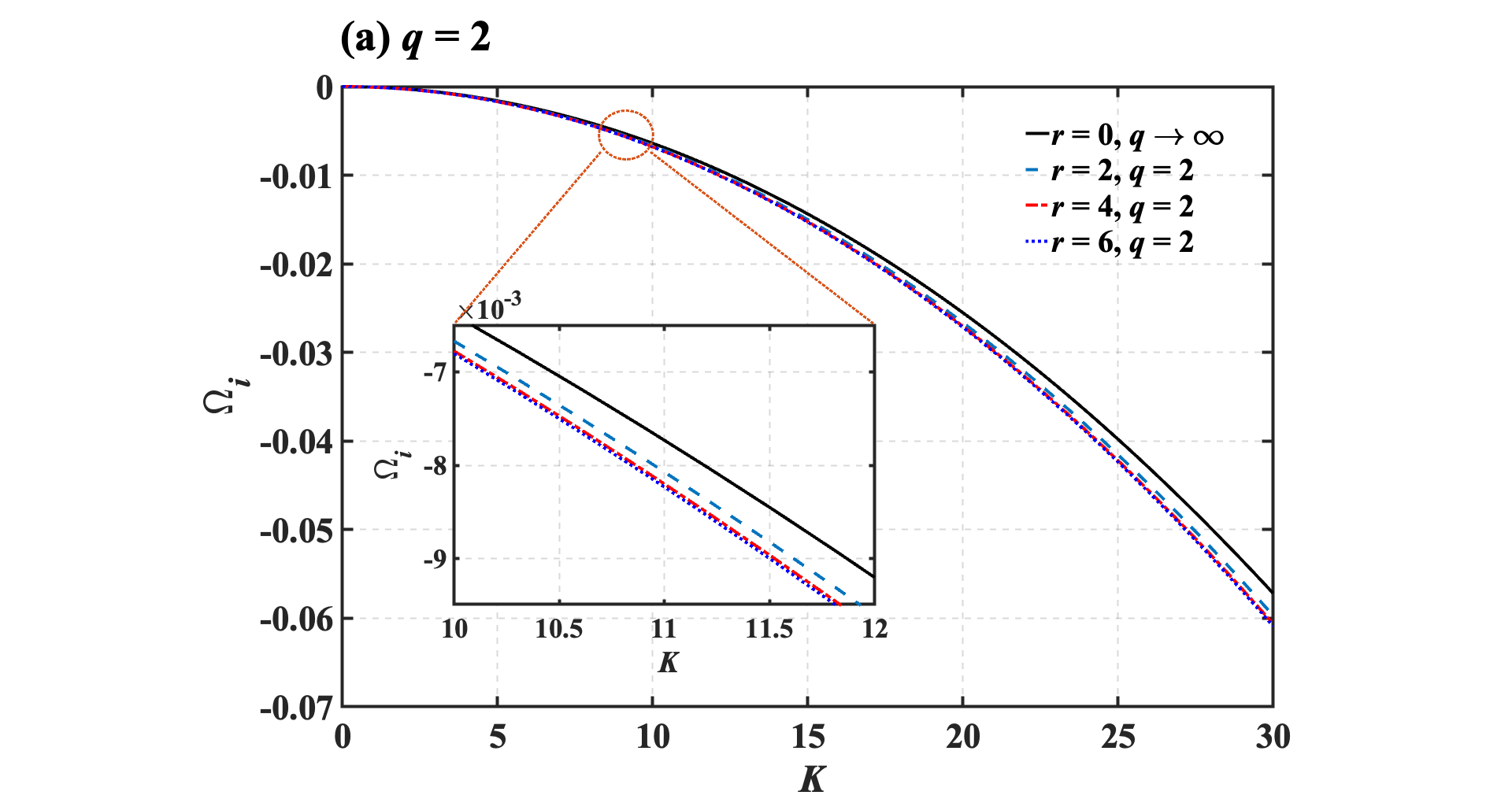}
    \end{subfigure}
    \hfill
    \begin{subfigure}{0.49\textwidth}
    \centering
    \includegraphics[width=1\textwidth]{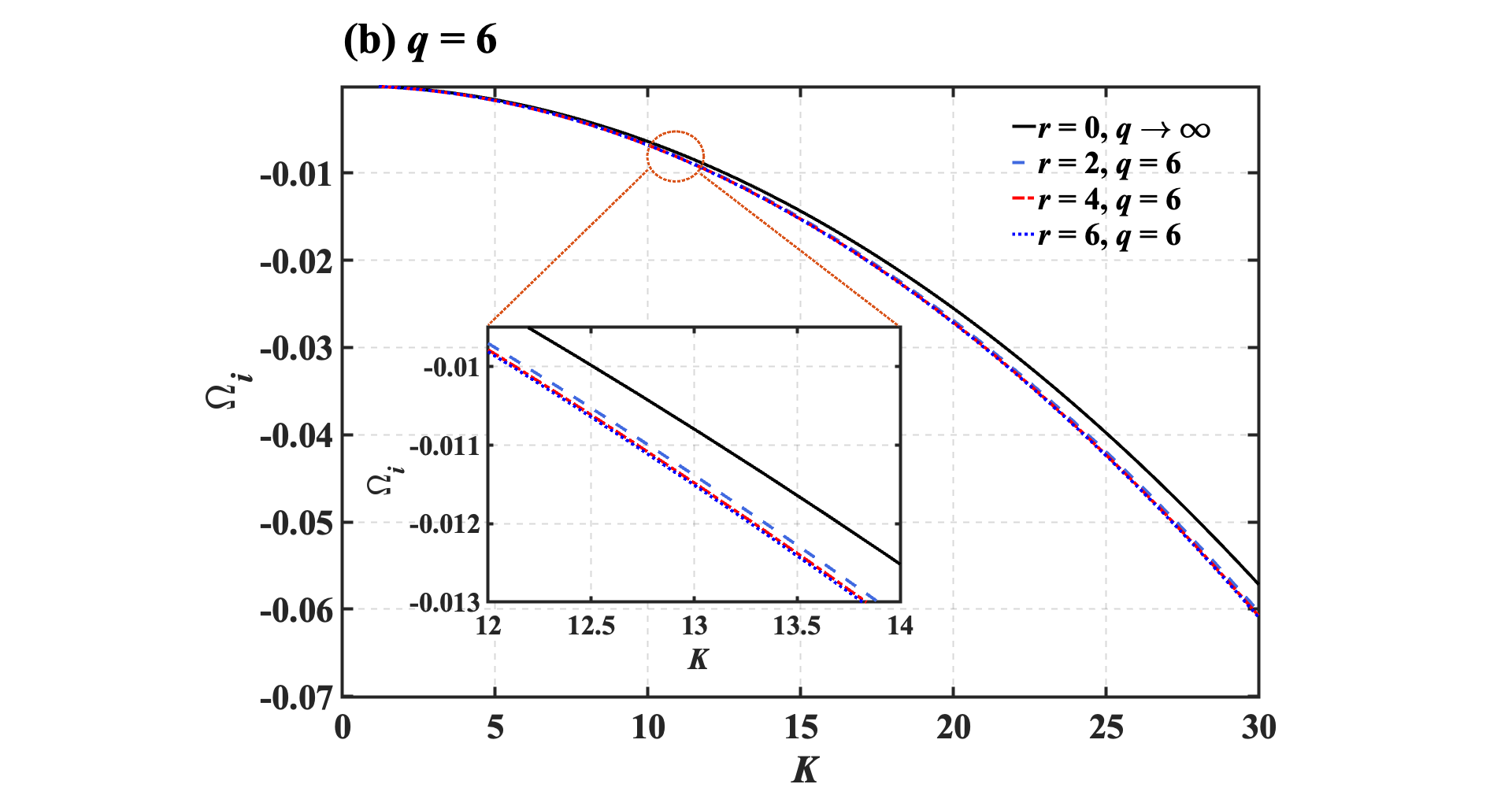}
    \end{subfigure}
    \caption{Same as figure \ref{fig:figure9}, but for different values of $r$ with (a) $q = 2$ and $q=6$. The various curves respectively link to: (a) $r = 0,\ q \to \infty$ (black solid line, taken as the Maxwellian reference line), (b) $r = 2$ (cyan dashed line), (c) $r = 4$ (red dashed-dotted line), and (d) $r = 6$ (blue dotted line).}
    \label{fig:figure10}
\end{figure}

A semi-analytic model is developed to assess the PMGC instability by incorporating the effects of the EiBI gravity, $(r,q)$-distributed electrons, and dust-polarization force. The derived quartic polynomial dispersion relation \eqref{eq:3.15} serves as an investigative tool for the oscillatory and propagatory dynamics associated with the PMGC instability. As detailed in section \ref{sec:Linearization and Fourier analysis}, the derived quartic dispersion relation \eqref{eq:3.15} is reduced to a linear dispersion relation using the approximation of ultra low-frequency (ULF) fluctuations. Finally, \eqref{eq:3.23} is utilized for further numerical analysis, and the resulting graphical representations (figures \ref{fig:figure2}-\ref{fig:figure10}) are respectively interpreted. Various input values of significant physical parameters are adopted from reliable sources in the literature. It includes $m_e = 9.1 \times 10^{-28}$ \unit{g}, $m_i = 1.6 \times 10^{-24}$ \unit{g}, $m_d = 4 \times 10^{-12}$ \unit{g}, $e = 4.80 \times 10^{-10}$ \unit{esu}, $q_d = - 50e$, $r_d = 1.5 \times 10^{-4}$ \unit{cm}, $n_{e0}= 1.2 \times 10^6$ \unit{cm^{-3}}, $n_{i0} = 4.95 \times 10^6$ \unit{cm^{-3}}, $n_{dc0} = 2$ \unit{cm^{-3}}, $n_{dn0} = 4$ \unit{cm^{-3}}, $T_e = 1$ \unit{eV}, $T_i = 0.8$ \unit{eV}, and $T_d = 0.01$ \unit{eV} \cite{dutta2016stability,dolai2020effects,Singha_2019,shukla2006jeans}. Accordingly, the Coulomb coupling parameter $\Gamma_\text{Cou}$ is estimated as $\sim 0.011 (\ll 1)$, which indicates a weakly coupled DMC. Additionally, the Jeans angular frequency $(\omega_J)$ is calculated as $4.49 \times 10^{-9}$ \unit{rad . s^{-1}}, the dust thermal velocity $(V_\text{td})$ is found as $8.9 \times 10^{-2}$ \unit{cm . s^{-1}}, and the dust-plasma oscillation frequency $(\omega_{pd})$ is determined as $1.04$ \unit{rad . s^{-1}}. Furthermore, the effective dust-plasma Debye length for $(r,q)$-distribution $(\lambda_{D_{r,q}})$ is obtained as $2.65 \times 10^{-1} $ \unit{cm}.

\par
As in figure \ref{fig:figure2}, we illustrate the impact of the polarization force on the PMGC instability in the presence of $(r,q)$-distributed electrons within the considered EiBI gravity framework. Here, the Jeans-scaled growth rate $(\Omega_i)$ of the PMGC instability is plotted against the Jeans-scaled angular wavenumber $(K)$ for different values of polarization parameter $(R=0,0.2,0.4,1.2,\text{ and }1.6)$. The EiBI parameter is kept fixed at $\chi=2 \times 10^6$ \unit{g^{-1}.cm^5. s^{-2}}, the Jeans-scaled radial distance at $\xi=30$, and the multi-order non-thermality coefficient at $A=1.4$ (for $r=0$ and $q=5)$. It demonstrates that for $R < 1$, the attenuation of the PMGC instability at shorter wavelengths (acoustic-like) is significantly more prominent compared to the longer wavelengths (gravitational-like), whereas for $R > 1$, the growth of the instability is more significant for shorter wavelengths (acoustic-like), i.e., larger $K$-values. It is clear from figure \ref{fig:figure2} that an increase in the value of $R$ reduces the damping nature of the PMGC instability, leading to a less stable system (most stable for $R=0$ and least stable for $R=1.6$). This indicates that within the EiBI gravity framework, the polarization force serves as a destabilizing factor for negatively charged dust. Previous studies have also reported a similar destabilizing nature of the polarization force on gravitational collapse due to the Jeans instability in simplified model configurations \cite{dolai2020effects,dutta2016stability,prajapati2011effect}. It should be emphasized that, for the weakly coupled DMC under consideration, with the specified physical parametric values, the polarization interaction parameter is calculated to be $R \sim 1.52 \times 10^{-5}$. Therefore, the polarization force $(\Vec{F}_p)$ is not expected to significantly influence the stability of the system.

\par 
Figure \ref{fig:figure3} is same as figure \ref{fig:figure2}, but portrayed in a wave space defined by Jeans-scaled radial distance $(\xi)$ and Jeans-scaled angular wavenumber $(K)$. In this scale of $\xi$, no significant effect over the PMGC fluctuation is observed; however, changes in $R$-values distinctly affect the $\Omega_i$-value. At $\xi=30$, the Jeans-scaled imaginary angular frequency has a minimum value $(\Omega_i=-7.6 \times 10^{-2})$ for $R=0$ and a maximum value $(\Omega_i = 4.07 \times 10^{-2})$ for $R=1.6$, suggesting the growth of the instability escalates with $R$. This supports the findings from figure \ref{fig:figure2}. The destabilizing effect of the polarization force and its underlying physics can be deduced from the effective electrostatic force $(\Vec{F}_\text{eff})$ acting on the dust grains. The effective electrostatic force acting on the dust grains subjected to an external electric field $\Vec{E}$ is given by $\Vec{F}_\text{eff} = (\Vec{F}_e+\Vec{F}_p) = q\Vec{E} -q_d^2\nabla\lambda_D /(2\lambda_D^2)$ \cite{hamaguchi1994polarization,khrapak2009influence}. Typically, the polarization force has a minimal impact on system stability. However, it becomes significant as the dust grain size increases \cite{asaduzzaman2012roles,khrapak2009influence,dolai2020effects}. Large-sized dust grains tend to have higher charges $(q_d)$, leading to an increased $R$-value (since $R \propto |q_d|$). Thus, with the rise of the dust-polarization force, the counteracting electrostatic force that opposes the inward gravitational pull weakens. Consequently, the gravitational force becomes predominant over the electrostatic force, leading to system instability. When $R \ge 1$, the dust polarization force exceeds the corresponding electrostatic force. It causes the effective electrostatic force $(\Vec{F}_\text{eff})$ to fail as a stabilizing force against the gravitational force \cite{chen2016nonlinear}, ultimately triggering the self-gravitational collapse of the interstellar cloud.

\par
In a similar way, figure \ref{fig:figure4} depicts a color-spectral top-view projection of figure \ref{fig:figure3}. This graphic projection is utilized for a more detailed examination of the conjugated relationship of $\Omega_i$ with $\xi$. Interestingly, it is found that in the vicinity of the central region of the DMC $(\xi \sim 0)$, the variation of $\Omega_i$ seems to vanish abruptly. This modal behaviour indicates the possible signatures of a microphysical correlation between $\Omega_i$ and $\xi$ depictable with $\Omega_i = f(\xi)$.

\par
Figure \ref{fig:figure5} displays similar features as figure \ref{fig:figure3}, but shown in a magnified view of $\xi$ and an extended range of $K$ (encompassing the intervals $\xi=0-3$ and $K=0-50$). Apart from the polarization effects noticed in figures \ref{fig:figure2} - \ref{fig:figure4}, a distinct spatial variation in $\Omega_i$ is observed in the low-$\xi$ region $(\sim 0-0.2)$. It indicates that in the proximity of the spherical DMC core, the PMGC instability shows a rapidly escalating (diminishing) tendency for $R < 1$ ($R \ge 1$). However, beyond $\xi\sim 0.2$, the instability reaches a plateau and spreads uniformly across the system. This plateau region can be interpreted as the result of the interaction between gravity and effective electrostatic force. As $\xi$ continues to increase, the effective electrostatic force within the cloud becomes adequate to counteract the gravitational collapse of the DMC.
\par
Figure \ref{fig:figure6} illustrates the EiBI gravity effect on the PMGC instability in the presence of polarization force and $(r,q)$-distributed electrons. For any self-gravitationally bounded astrophysical object of size $L$, the value of $\chi$ should be less than $GL^2$, $G$ being the universal gravitational constant \cite{avelino2012eddington}. For our system having modified critical Jeans length, $\lambda_J\sim 1.37\times10^8$ \unit{cm}, the constraint is found to be $\chi<1.33\times10^9$ \unit{g^{-1}.cm^5.s^{-2}}. The $\chi$-value is varied in the range from $-4 \times 10^6\ \unit{g^{-1}.cm^5.s^{-2}} \text{ to } 4\times 10^6$ \unit{g^{-1}.cm^5.s^{-2}}. It is noteworthy here that, $\chi=0$ serves as the Newtonian reference in the non-local gravity, which is in fact, a crucial feature for assessing the observed trends. All the other parameters are kept the same as mentioned at the beginning of this section, except $R=0.2$ and $A=1.4\ (r=0,q=5)$.

\par
It is clear from figure \ref{fig:figure6} that an elevation in the negative EiBI parameter $(\chi)$ results in the deceleration of the PMGC instability, thereby enhancing the stability of the system in comparison to those in the Newtonian scenarios. Conversely, an increase in the positive EiBI parameter $(\chi)$ drives the system towards a state of higher instability. Therefore, it can be deduced that the positive EiBI parameter functions as a destabilizing factor, while the negative EiBI parameter acts as a stabilizing factor for the system being analyzed. Previous literature has also reported a similar impact of the EiBI gravity on the Jeans instability \cite{yang2020jeans}.

\par
Figure \ref{fig:figure7} depicts similar characteristics to those in figure \ref{fig:figure6}. In this representation, the spatial aspects of the PMGC instability are illustrated by jointly varying the Jeans-scaled radial distance $(\xi)$ and Jeans-scaled wavenumber $(K)$. Here, only two values of the EiBI parameter ($\chi=\pm\ 2\times10^6$ \unit{g^{-1}.cm^5.s^{-2}}) are examined alongside the Newtonian case $(\chi=0)$. All other pertinent physical parameters remain unchanged from those utilized in figure \ref{fig:figure6}. Similar to the findings in figure \ref{fig:figure6}, it is noted that a positive (negative) EiBI parameter $(\chi)$ acts as a destabilizing (stabilizing) factor in the presence of non-thermal $(r,q)$-distributed electrons and sheath-polarization of the Debye sheaths around the dust grains present in the DMC. Additionally, a color-spectral top-view projection of the same is shown in figure \ref{fig:figure8}.

\par
Figure \ref{fig:figure9} elucidates the influence of $(r,q)$-distributed electrons on the PMGC instability within the EiBI gravity framework, taking the polarization force into account. The variation of Jeans-scaled imaginary angular frequency $(\Omega_i)$ with the Jeans-scaled angular wavenumber $(K)$ is depicted for different values of $q\ (=3,5,10)$ with (a) $r=0$ and (b) $r=1$. In the limiting condition $r=0$ and $q \to \infty$, the value of $A$ converges to $1$. Hence, the line corresponding to $r=0$ and $q \to \infty$ serves as the Maxwellian reference. This numerical analysis maintains consistent fixed equilibrium values as used before except $R=0.2$ and $\chi=2\times10^6$ \unit{g^{-1}.cm^5.s^{-2}}.
\par
It is observed in figure \ref{fig:figure9}a that for $r=0$, the different curves exhibit a compelling propensity to converge towards the Maxwellian reference as the value of $q$ increases. This outcome is completely inline with the previous prediction found on the $(r,q)$-distribution laws. Interestingly, the results diverge significantly from the previous case when $r = 1$. It is seen in figure \ref{fig:figure9}b that the $q$-value increments deviate the curves away from the Maxwellian reference, indicating the emergence of non-Maxwellian behavior. Moreover, it is evident from figure \ref{fig:figure9} that $q$-value increments dampen the instability of the system for both the cases $(r=0,1)$. Nevertheless, for $r=0$, the Maxwellian system is observed to be more stable than the non-Maxwellian system, whereas for $r=1$, the non-Maxwellian system is more stable.

\par
In a similar way, in figure \ref{fig:figure10}, we portray the same as figure \ref{fig:figure9}, but for varying values of $r\ (=2,4,6)$ with (a) $q=2$ and (b) $q=6$. As figure \ref{fig:figure9}, it is evident that an increase in the $r$ spectral parameter enhances the system stability accompanied by a rise in the non-Maxwellian (non-thermal) characteristics. A comparative analysis of figures \ref{fig:figure9}-\ref{fig:figure10} reveals a consistent trend. Both the non-thermal spectral parameters ($r$ and $q$) play a significant role in influencing the system stability and non-Maxwellian characteristics. While elevating the $q$ value consistently improves stability across different $r$ values and vice versa, it also steers the system away from the Maxwellian velocity distribution. Additionally, it is notable that at longer wavelengths (gravitational-like), the system maintains closer proximity to the Maxwellian characteristics, whereas at shorter wavelengths (acoustic-like), the deviation from thermal behaviour becomes more pronounced. The stability of the non-Maxwellian (non-thermal) systems can be comprehended by noting that as the electron energy increases, their mobility also rises and vice versa. Consequently, there is a rapid accumulation of electrons at the dust grain surfaces, leading to an escalation in the dust charge. The enhanced dust charge at the cost of thermal loss of electrons results in an increased electrostatic repulsion among the dust grains, counterbalancing the gravitational forces acting inward.

\section{Conclusions}\label{sec:Conclusions}
The proposed study explores the collective impact of the EiBI gravity, $(r,q)$-distributed electrons, and dust-polarization force on the pulsational mode of gravitational collapse (PMGC) in complex spherical dust molecular cloud (DMC) fluids. Application of a standard spherical normal mode analysis in the EiBI-modified DMC results in a unique form of a generalized linear quartic dispersion relation. The derived linear dispersion relation, if all the newly added astronomical sophistications are switched off, fairly corroborates with the previous results reported in the literature \cite{dwivedi1999pulsational}. After its necessary validation checkup, the derived PMGC dispersion relation is numerically illustrated with judicious parametric inputs to explore diverse PMGC stability features in different realistic astronomical circumstances.
\par 
The PMGC instability is found to be significantly more pronounced for longer wavelengths (gravitational-like) than the shorter wavelengths (acoustic-like). Additionally, the instability exhibits a rapidly growing tendency in the vicinity of the DMC core. It shows a saturating propensity from the cloud center outwards. It is seen that the polarization force and the positive EiBI parameter act as destabilizing factors, while the negative EiBI parameter acts in favour of the DMC stability. The non-thermal $(r,q)$-distributed electrons produce a DMC stabilizing influence. It is important to note that the deviation from the Maxwellian (thermal) characteristics is significant at shorter wavelengths (acoustic-like), whereas at longer wavelengths (gravitation-like), there is no substantial distinction between the Maxwellian (thermal) and non-Maxwellian (non-thermal) features in the thermo-statistical perspective. However, the non-Maxwellian systems are found to be more stable than the Maxwellian ones.
\par
The outcomes of the PMGC instability analyses presented in this work could be utilized to enhance the understanding the diverse processes of astrophysical structure formation in the H \textsc{ii} regions (ionized hydrogen regions) of the molecular clouds, such as Sh2-87 \cite{biswas2024star}, Sh2-88B \cite{deharveng2000stellar}, Sh2-235 \cite{kirsanova20203d}, etc. The Jeans-like instability in the EiBI gravity framework discussed in this study may lead to the fragmentation of extremely dense interstellar DMCs into smaller substructures that cannot withstand their own gravitational forces. These results may be helpful in further probing the EiBI gravity role in various other similar structures. It may include various star forming molecular cloud zones in our galaxy, such as the Central Molecular Zone \cite{lu2020alma}, the Orion molecular cloud \cite{palau2018thermal}, etc. While there have not been any substantial observational studies of the PMGC model reported yet, the recently launched JWST enables us to observe star-forming regions beyond the Milky Way and its satellite galaxies \cite{peltonen2024jwst}. Already, the JWST has uncovered galaxies with $z>10$ ($z$ being the redshift). Additionally, the Atacama Large Millimeter Array (ALMA) facilitates the observation and analysis of gas and star formation process in galaxies with $z > 4$, on scales less than a kiloparsec \cite{FREUNDLICH2024100059}. These developments could open up new avenues for future investigations in light of the real-time insightful observations in the relevant PMGC research areas.
\par
It is important to note here that the effects, like viscoelasticity (in a strongly coupled system), dust charge fluctuation, ion drag, turbulence, and magnetic field are not considered in this work. Additionally, there are numerous other extensions of general relativity as well as several other velocity distributions, the effects of which on the behaviour of instabilities are yet to be explored. Therefore, it is conclusively admitted that there is potential scope for future improvements to the presented model by incorporating such important plasma-fluidic and thermo-statistical features of realistic astronomical significance.

\acknowledgments
The Department of Physics at Tezpur University is gratefully acknowledged for its invaluable cooperation. Special appreciation is due to all the members of the Astrophysical Plasma and Nonlinear Dynamics Research Laboratory (APNDRL) for their collaborative efforts and insightful discussions, which have enriched the intellectual environment of our research pursuits. Special acknowledgment is given to Mr. Souvik Das of the Department of Physics, Tezpur University, for his meticulous attention to detail and unwavering support in manuscript formatting. Finally, Mr. Siddharth Saikia, affiliated to the University of New South Wales, is also hereby acknowledged for his help in writing the elementary parts of the manuscript. 

\appendix
\section{Polarization force for negatively charged dust}\label{sec:appendix}
The expression for the polarization force \cite{khrapak2009influence,hamaguchi1994polarization} in a complex (dusty) plasma system can be expressed as  
\begin{equation}\label{eq:5.1}
    F_p = - \left(\frac{q_d^2}{2}\right)\left(\frac{\nabla\lambda_D}{\lambda_D^2}\right);
\end{equation}
where, 
\begin{equation}\label{eq:5.2}
    \lambda_D=\lambda_{Di}\left(1+\frac{\lambda_{Di}^2}{\lambda_{De}^2}\right)^{-1/2}=\lambda_{Di}\left(\frac{n_i T_e}{n_i T_e + n_e T_i}\right)^{1/2}.
\end{equation}
 Here, $\lambda_{De(i)} = \left(k_B T_{e(i)} / 4\pi e^2 n_{e(i)}\right)^{1/2}$ represents the electron (ion) Debye length, with $T_{e(i)}$ being the electron (ion) temperature (in $K$), $e$ the elementary charge (electronic), and $n_{e(i)}$ corresponding to the electron (ion) number density. For a dusty plasma comprising of negatively charged dust grains, $n_i \gg n_e$ and $T_e \gg T_i$. Therefore, we can judiciously use the approximation $n_eT_i \ll n_iT_e$ in \eqref{eq:5.2} to get

\begin{equation}\label{eq:5.3}
    \lambda_D=\left(\frac{k_BT_i}{4\pi e^2 n_i}\right)^{1/2} \approx \lambda_{Di}\hspace{1mm}.
\end{equation}
Now, taking gradient on both sides of \eqref{eq:5.3} yields 

\begin{equation}\label{eq:5.4}
    \begin{aligned}
        \nabla\lambda_D=\nabla\lambda_{Di}=\nabla\left(\frac{k_BT_i}{4\pi e^2 n_i}\right)^{1/2} = -\frac{1}{2}\left(\frac{k_BT_i}{4\pi e^2 n_i}\right)^{1/2}\left(\frac{\nabla n_i}{n_i}\right).
    \end{aligned}
\end{equation}
Dividing both sides of \eqref{eq:5.4} by $\lambda_D^2$, one obtains
\begin{equation}\label{eq:5.5}
    \begin{aligned}
        \frac{\nabla\lambda_D}{\lambda_D^2}=-\frac{1}{2\lambda_D}\left(\frac{\nabla n_i}{n_i}\right).
    \end{aligned}
\end{equation}
Considering the ions to be Maxwellian, we can use $T_i\nabla n_i = - n_i e \nabla \phi$ in \eqref{eq:5.5} as
\begin{equation}\label{eq:5.6}
     \frac{\nabla\lambda_D}{\lambda_D^2}=-\frac{1}{2\lambda_D}\left(\frac{-n_ie\nabla\phi}{n_iT_i}\right)=\frac{1}{2\lambda_D}\left(\frac{e\nabla\phi}{T_i}\right).
\end{equation}
Finally, substituting $\nabla\lambda_D / \lambda_D^2$ in \eqref{eq:5.1}, one gets
\begin{equation}\label{eq:5.7}
    \begin{aligned}
    F_p &= - \left(\frac{q_d^2}{4\lambda_D}\right)\left(\frac{e\nabla\phi}{T_i}\right) = -|q_d| \left(\frac{ |q_d| e}{4T_i\lambda_{Di0}}\right)\left(\frac{n_i}{n_{i0}}\right)^\frac{1}{2} \nabla\phi = -|q_d| R \left(\frac{n_i}{n_{i0}}\right)^\frac{1}{2} \nabla\phi,
    \end{aligned}
\end{equation}
where, $R= \left(|q_d| e/4T_i\lambda_{Di0}\right)$ is known as the polarization interaction parameter with $\lambda_{Di0} = \left(k_B T_{i} / 4\pi e^2 n_{i0}\right)^{1/2}$ being the equilibrium ion Debye length. 
\par
It is evident from \eqref{eq:5.7} that the polarization force is directly proportional to the magnitude of dust charge $(|q_d|)$, polarization parameter $(R)$, and the gradient of electrostatic potential $(\nabla \phi)$ in a plasma system.



\bibliographystyle{JHEP}
\bibliography{ms}
\end {document}